\documentclass[aps,prd,preprint,eqsecnum,preprintnumbers,nofootinbib,byrevtex,showpacs,showkeys,groupedaddress,amssymb,amsmath,floatfix]{revtex4}
\usepackage{bm}
\usepackage{graphics}
\usepackage{graphicx}
\usepackage{epsfig}
\usepackage{booktabs,multirow}
\usepackage{hhline}
\usepackage{slashed}
\usepackage{rccol}
\usepackage{dcolumn}
\usepackage{ mathrsfs }
\usepackage[capitalize]{cleveref}

\begin{document}
\title{Single Neutralino Production at the LHC}
\author{A.~I.~Ahmadov$^{1,2}$}
\email{E-mail:ahmadovazar@yahoo.com}
\author{M.~Demirci$^{1}$}
\email{E-mail:phy_mdemirci@yahoo.com}
\affiliation{$^{1}$Department of Physics, Karadeniz Technical University, 61080 Trabzon, Turkey \\
$^{2}$Institute for Physical Problems, Baku State University, Z.
Khalilov st. 23, AZ-1148, Baku, Azerbaijan}%

\date{\today}

\begin{abstract}
We consider that the direct production of a single neutralino in
proton-proton collision at the CERN Large Hadron Collider focusing
on the lightest neutralino is possibly a candidate for the dark
matter and escapes detection. We present a comprehensive
investigation of the dependence of total cross sections of the
processes $pp(q\overline
q)\rightarrow\widetilde\chi_{i}^{0}\widetilde{\text{g}}$, $pp(q
\text{g}) \rightarrow\widetilde\chi_{i}^{0}\widetilde{q}_{L,R}$,
$pp(q\overline{
q}^{\prime})\rightarrow\widetilde\chi_{i}^{0}\widetilde\chi_{j}^{+}$
at tree-level and
$pp(\text{g}\text{g})\rightarrow\widetilde\chi_{i}^{0}\widetilde{\text{g}}$
at one-loop level, on the center-of-mass energy, on the $M_2$-$\mu$
mass plane, on the squark mass and on the $\tan\beta$ for the Constrained Minimal Supersymmetric Standard Model and the three extremely different scenarios in
the Minimal Supersymmetric Standard Model. In particular, the cross section of the process $p p \to
\widetilde\chi_{2}^{0}\widetilde\chi_{1}^{+}$ in the gaugino-like
scenario can reach about 0.6 (1.7) pb at a center-of-mass energy of
$\sqrt{s}=$ 7 (14) TeV. We derive therefrom that our results might
lead to new aspects corresponding to experimental explorations, and
these dependencies might be used as bases of experimental research
of the single neutralino production at hadron colliders.
\end{abstract}

\maketitle

\section{\bf Introduction}
The supersymmetric (SUSY) theories \cite{1971} have long been one of
the leading candidates for new physics beyond the Standard model
(SM). They postulate the existence of SUSY particles (sparticles)
whose spin differ by one-half unit with respect to that of their SM
partner \cite{Haber,Nilles,Kazakov}, and introducing these new
particles provides solutions to the hierarchy problem. The existence
of these supersymmetric particles can be determined at the Large
Hadron Collider (LHC) and the International Linear Collider (ILC),
who might supply the experimental facilities. However, even if
most of the supersymmetric particles are produced at colliders, they will not
be detected because they will eventually decay into the
lightest supersymmetric particle (LSP) on condition that the
R-parity \cite{Fayet,Martin} is conserved. As consequences of the
R-parity conservation, sparticles can only be created (or destroyed)
in pairs and the LSP is absolutely stable, which is generally
assumed to be a weakly interacting massive particle, and so making
it an excellent candidate for astrophysical dark matter
\cite{DarkMatter,DarkMatter2} that is one of the attractive features of
SUSY. In the majority of SUSY breaking models, the LSP is the
lightest neutralino, and it occurs at the end of the decay chain of each
supersymmetric particle. For these purposes, a detailed study of the
lightest neutralino is of great importance for the theoretical and
phenomenological aspects of SUSY.

Among all the supersymmetric models, the Minimal Supersymmetric
Standard Model (MSSM), which is almost the direct version of the
supersymmetric of the SM, has an extra Higgs doublet and general
SUSY breaking soft terms. The superpartners of the Higgs doublets
(Higgsinos) mix with the superpartners of the gauge bosons
(gauginos) to form four Majorana mass eigenstates called neutralino
$\widetilde{\chi}_{i}^{0}$, $i=1,2,3,4$ and two charged mass
eigenstates called charginos $\widetilde{\chi}_{j}^{\pm}$, $j=1,2$
in the MSSM. The gaugino-Higgsino
decomposition of the neutralinos and charginos includes significant
knowledge regarding the mechanism of the supersymmetry breaking and
plays an essential role in the establishing the relic density of
the dark matter \cite{DarkMatter3}.

The experimental explorations of the supersymmetric particles are among the main tasks
of the experimental program at hadron colliders, especially at the LHC. Up to now, the ATLAS and CMS collaborations have chiefly concentrated on seeking the production of the strongly interacting squarks and gluinos. As a result, bounds on the masses of the squarks and gluinos are pushed to higher scales~\cite{ATLAS,CMS}, and the experimental attention starts to go towards the production of the electroweak slepton, neutralino and chargino.

The lower limit on the lightest neutralino mass ($m_{\widetilde{\chi}_{1}^{0}}$) is given about 46 GeV at 95$\%$ confidence level,  which can be obtained from the experimental bound on chargino mass in the MSSM at the large electron positron \cite{Abdallah}. However, this limit increases to well above 100 GeV from the strong restrictions set by the recent LHC data in the framework of the constrained MSSM (CMSSM) containing both gaugino and sfermion mass unification~\cite{PDG}.

In this paper, taking into account the allowed parameter space of
the MSSM, we present numerical results for the single neutralino
production processes in proton-proton collision at the LHC including a
neutralino in the final state as follows: the associated
subprocesses $q\overline
q\rightarrow\widetilde\chi_{i}^{0}\widetilde{\text{g}}$, $q \text{g}
\rightarrow\widetilde\chi_{i}^{0}\widetilde{q}_{L,R}$, $q\overline{
q}^{\prime}\rightarrow\widetilde\chi_{i}^{0}\widetilde\chi_{j}^{+}$
at tree-level and
$\text{g}\text{g}\rightarrow\widetilde\chi_{i}^{0}\widetilde{\text{g}}$
at one-loop level. There have been many works devoted to the study
of these processes in literature. For example, Refs.~\cite{Baer,Berger} focus on gluon/squark
produced in association with charginos and neutralinos at proton-proton
collision; Ref.~\cite{Allanach} discusses the feasibility of SUSY
monojet production at the LHC for measuring the
neutralino-squark-quark coupling; the automized
the next-to-leading-order QCD and SUSY-QCD corrections to the
squark-neutralino production are the focus of Ref.~\cite{Binoth}; Refs.~\cite{Beenakker,Debove} search for
associated production of charginos and neutralinos, and Ref.~\cite{Gounaris} focuses on single neutralino production.

One of the important approaches of our scenario consists of the
mechanism of choosing the input parameters. Unlike the above works, in our study we
have recovered the SUSY Lagrangian parameters as direct analytical
expressions of suitable physical masses without constraining any of them in
the MSSM so that we have principally concentrated on the
algebraically nontrivial inversion for the gaugino mass parameters. In other words,
using two chargino masses and $tan\beta$ as input parameters,
we obtain the other parameters, which are gaugino/Higgsino mass
parameters, the masses and mixing matrix of neutralino as outputs.

The neutralino mass eigenstates $\widetilde\chi_i^{0}$ (i=1,..,4)
are the linear superposition of the gauginos $\widetilde B$,
$\widetilde W^{3}$ and the Higgsinos $\widetilde H_1^{0}$,
$\widetilde H_2^{0}$ in the MSSM. In our case, the relative
importance of the production mechanisms ($\widetilde q_{L,R}$ and
$W^{+}$) depends on the strengths of the  $\widetilde B$, $\widetilde
W^{3}$ and the Higgsinos $\widetilde H_1^{0}$, $\widetilde H_2^{0}$
components of the $\widetilde\chi_{i}^{0}$; thus, significant
differences in the cross sections are to be expected for the case of
a gaugino-like, higgsino-like and mixing neutralino,
respectively. As we know, a supersymmetric neutralino is the
standard candidate for weakly interacting massive particles dark
matter. Despite this, it is still an open problem in SUSY.

By taking this information into account, it may be argued that the
calculation and analysis of the single neutralino production at
proton-proton collisions within the chosen scenarios is significant
from both theoretical and experimental points of view. Accordingly,
we investigate the dependence of total cross sections of
the single neutralino production processes on the center-of-mass energy, the $M_2$-$\mu$
mass plane, the squark mass and the $\tan\beta$ for CMSSM, and the three extremely different scenarios in
the MSSM.

The layout of the paper is as follows: In Section \ref{cs}, we provide analytical
expressions of the amplitudes and the cross sections of the relevant
subprocesses and also the corresponding couplings. In Section
\ref{results}, we present detailed numerical results of the cross
sections for each scenario and discuss the dependence of the cross
section on the MSSM model parameters. Our conclusions are
presented in Section~\ref{Conc}. Finally,
we summarize general information about the neutralino/chargino
sector in the MSSM and present formulas related to obtaining
neutralino masses in Appendix~\ref{Appendix}.

\section{Analytic Results of the Cross Sections for The Single Neutralino Production}\label{cs}
In this section, we present succinct definitions of generalized
corresponding to couplings in SUSY, and analytically expressions of
the relating partonic cross sections for single neutralino
production. Supplemental information about neutralinos at
proton-proton collisions can be acquired from the single neutralino
production triggered by the following subprocesses:
\begin{eqnarray} \label{eq:absinglen}
a^{k}(p_1)~b^{l}(p_2)\rightarrow
\begin{cases}
&\widetilde\chi_{i}^{0}(k_1,E_1,m_i)~\widetilde{\text{g}}^{m}(k_2,E_2,m_j) \\
&\widetilde\chi_{i}^{0}(k_1,E_1,m_i)~\widetilde q_{L,R}^{m}(k_2,E_2,m_j)\\
&\widetilde\chi_{i}^{0}(k_1,E_1,m_i)~\widetilde\chi_{j}^{\pm}(k_2,E_2,m_j) \\
\end{cases}
\end{eqnarray}
These are presented for the initial partons $a,b = q,
\overline{q},\text{g}$ whose masses can be neglected. Here, $p_{1}$ and
$p_{2}$ denote the four-momentum of the initial partons, $k_1$ and
$k_2$ represent the four-momentum of the two final states of a neutralino together with
a gluino (or squark or chargino), respectively. We represent by
$k,l,m$ the color indices for the corresponding particles.

The Mandelstam variables for subprocesses~\eqref{eq:absinglen} are
given as
\begin{equation}
\hat s=(p_1+p_2)^2, \quad \hat t=(p_1-k_1)^2,\quad \hat
u=(p_1-k_2)^2.
\end{equation}
Denoting by ($p, \theta$) the momentum and scattering angle in the
center-of-mass frame of the final states, we get center-of-mass
energy and momentums as follows:
\begin{equation} \label{eq:cm}
\begin{split}
& p=\frac{1}{2 \sqrt{\hat s}}\sqrt{(\hat s-m_i^2-m_j^2)^2-4m_i^2
m_j^2},\\
& E_1=\frac{\hat s+m_i^2-m_j^2}{2 \sqrt{\hat s}},~E_2=\frac{\hat
s+m_j^2-m_i^2}{2 \sqrt{\hat s}},\\
& p_1=\frac{\sqrt{\hat s}}{2}(1,0,0,1),~p_2=\frac{\sqrt{\hat
s}}{2}(1,0,0,-1),\\
& k_1=(E_1,p \sin\theta,0,p \cos\theta),~k_2=(E_2,-p \sin\theta,0,-p
\cos\theta).
\end{split}
\end{equation}

We give generalized electroweak couplings for the corresponding
single neutralino production in the MSSM. The square of the weak coupling constant
$g^2=e^2/\sin^2\theta_W$ is defined in terms of the electromagnetic fine
structure constant $\alpha=e^2/4 \pi$ and electroweak mixing angle
$c_W=\cos\theta_W, s_W=\sin\theta_W$. Following the standard
notation, the $W$-chargino-neutralino interaction vertices are
proportional to couplings as follows \cite{Haber}:
\begin{equation} \label{eq:O_ij}
\begin{split}
& O_{ij}^L=-\frac{c_W}{\sqrt2} N_{i4}V_{j2}^* + c_W N_{i2}V_{j1}^*,\\
& O_{ij}^R=\frac{c_W}{\sqrt2} N_{i3}^*U_{j2} + c_W N_{i2}^* U_{j1}.\\
\end{split}
\end{equation}
We neglect masses of the initial partons and generational mixing in
the (s)quark sectors, the gaugino-squark-quark interaction vertices
are proportional to the corresponding neutralino-squark-quark
couplings \cite{Gunion,Rosiek},
\begin{equation} \label{eq:C_Nqsq}
\begin{split}
& C_{\widetilde \chi_i^0 \widetilde q q}^L=\bigl[(e_q-I_q^3)s_W N_{i1}+ I_q^3 c_W N_{i2}\bigr],\\
& C_{\widetilde \chi_i^0 \widetilde q q}^R=-e_q s_W N_{i1}^*,\\
\end{split}
\end{equation}
and the corresponding chargino-squark-quark couplings (for
$q,q^{\prime}=u,d$ quarks)
\begin{equation} \label{eq:C_Cqsq}
\begin{split}
& C_{\widetilde \chi_j^+ \widetilde q q^{\prime}}^L=\frac{c_W}{\sqrt
2}(U_{j1}~\delta_{q^{\prime}u} +
V_{j1}^*~\delta_{q^{\prime}d})-\frac{c_W(m_dU_{j2}~\delta_{q^{\prime}u}
+m_u V_{j2}^*~\delta_{q^{\prime}d})}{2m_W(\cos\beta~\delta_{q^{\prime}u}+\sin\beta ~\delta_{q^{\prime}d})},\\
& C_{\widetilde \chi_j^+ \widetilde q
q^{\prime}}^R=-\frac{m_{q^{\prime}}c_W(V_{j2}^*~\delta_{q^{\prime}u}
+U_{j2}~\delta_{q^{\prime}d})}{2m_W(\sin\beta~\delta_{q^{\prime}u}+\cos\beta ~\delta_{q^{\prime}d})},\\
\end{split}
\end{equation}
where $I_q^3$ is the weak isospin quantum number such that
$I_q^3=\pm 1/2$ for left-handed and $I_q^3=0$ for right-handed up-
and down-type quarks, $e_q$ denotes their fractional electromagnetic
charge, and the matrices $N,U$ and $V$ are neutralino and chargino
mixing matrices, respectively. The couplings of the neutralino to
quark, squark, chargino and $W$ boson are determined by the
corresponding elements of the mixing matrices ($N_{ij} , U_{ij}
,V_{ij}$), as shown in the above relations. The relevant couplings of the particles for single
neutralino production are derived from the following interaction
Lagrangians of the MSSM \cite{Haber}:
\begin {equation} \label{eq:L_singleN}
\begin{split}
& L_{\widetilde{\chi}_{i}^{0}\widetilde{q} q}= -\frac{\sqrt2 g}{c_W}
\overline{q}\left[C_{\widetilde \chi_i^0 \widetilde q q}^{L*}
P_{L}+C_{\widetilde \chi_i^0 \widetilde q
q}^{R*}P_{R}\right]\widetilde{\chi}_{i}^{0}\widetilde{q}_{L,R},\\
& L_{q\widetilde{q}\widetilde{\text{g}}}= -\sqrt2 g_s T_{jk}^a
\left[\overline{\widetilde{\text{g}}}_a P_{L} q^k
\widetilde{q}_L^{j*}+\overline{q}^j P_{R}\widetilde{\text{g}}_a
\widetilde{q}^k_L - \overline{\widetilde{\text{g}}}_a P_{R} q^k
\widetilde{q}_R^{j*}-\overline{q}^j P_{L}\widetilde{\text{g}}_a \widetilde{q}^k_R \right],\\
& L_{\widetilde{q}\widetilde{q}\text{g}}= -i g_s G_{\mu}^a\left[\widetilde{q}^{j*}T_{jk}^a  \partial^{\mu} \widetilde{q}^k- \partial^{\mu}\widetilde{q}^{j*} T_{jk}^a\widetilde{q}^k\right],\\
& L_{W^+\widetilde{\chi}_{i}^{0}\widetilde{\chi}_{j}^{+}}=
\frac{g}{c_W} W_\mu \overline{\widetilde{\chi}_{i}^{0}}
\gamma^{\mu} \left[O_{ij}^{L} P_{L}+O_{ij}^{R}P_{R}\right]\widetilde{\chi}_{j}^{+},\\
&L_{\widetilde{\chi}_{j}^{+}\widetilde{q} q^\prime}= -\frac{\sqrt2
g}{c_W} \overline{q}\left[C_{\widetilde \chi_j^+ \widetilde q
q^{\prime}}^{L*} P_{L}+C_{\widetilde \chi_j^+ \widetilde q
q^{\prime}}^{R*}P_{R}\right]\widetilde{\chi}_{j}^{+}\widetilde{q}_{L,R},\\
\end{split}
\end {equation}
where
$\widetilde{\chi}_{i}^{0},\widetilde{\chi}_{j}^{+},q,\widetilde{q}_{L,R}$
and $\widetilde{\text{g}}$ are four-component spinor fields,
$P_{R,L}=\frac{1}{2}(1\pm\gamma^5)$, $T_{jk}^a$ is a color generator,
and strong coupling $g_s=\sqrt{4\pi\alpha_s}$. The running strong
coupling constant $\alpha_s$ is given as follows:
\begin{equation} \label{eq:alfas}
\alpha_s(Q^2)=\frac{4\pi}{(11-\frac{2}{3}n_f) ln(Q^2/\Lambda^2)},
\end{equation}
where $\Lambda$ is the QCD scale parameter, and $n_f$ is the number
of active flavors at the energy scale $Q$ that can be chosen as the
average of the final particle masses.

The total cross sections for subprocesses can be obtained by using
the following formula~\cite{Greiner}:
\begin{equation} \label{eq:sigma}
\hat \sigma(\hat s)=\int_{\hat{t}^-}^{\hat{t}^+}d\hat t ~\frac{d\hat
\sigma}{d\hat t},
\end{equation}
where $\hat{t}^\pm=1/2\bigl[(m_i^2+m_j^2-\hat s)\pm\sqrt{(\hat
s-m_i^2-m_j^2)^2-4m_i^2 m_j^2}\bigr]$. With the results from Eq.
\eqref{eq:sigma} for the relevant subprocess, the total unpolarized
hadronic cross sections in proton-proton collisions at
center-of-mass energy can be calculated by
\begin{equation} \label{eq:totalsigma}
\sigma(s)=\int_{(m_i+m_j)^2/s}^{1}d\tau~
\frac{d\mathcal{L}^{pp}_{ab}}{d\tau}
~\hat{\sigma}(\text{\textit{subprocess, at}}~ \hat s=\tau s),
\end{equation}
with the parton luminosity
\begin{equation} \label{eq:pluminosity}
\frac{d\mathcal{L}^{pp}_{ab}}{d\tau}=\int_\tau^1
\frac{dx_1}{x_1}\frac{1}{1+\delta_{ab}}\biggl[G_{a/h_1}(x_1,\mu_F)G_{b/h_2}(\frac{\tau}{x_1},\mu_F)
+G_{b/h_1}(x_1,\mu_F)G_{a/h_2}(\frac{\tau}{x_1},\mu_F)\biggr],
\end{equation}
where $G_{a/h_1}$ and $G_{b/h_2}$ are universal parton densities of
the partons $a,b$ in the hadrons $h_1,h_2$, which depend on the
longitudinal momentum fractions of the two partons $x_1,x_2$
($\tau=x_1 x_2$) at the unphysical factorization scale $\mu_F$. We
fix the factorization scale to the average mass of the final state
particles, $\mu_F=(m_i+m_j)/2$.

Considering each subprocess separately, we now present analytic
expressions of the amplitudes and the differential cross sections
for the single neutralino production in the following subsections.

\subsection{The subprocess $\boldsymbol{q\overline
q\rightarrow\widetilde\chi_{i}^{0}\widetilde{\text{g}}}$}
\begin{figure}[ht]
\includegraphics[width =12cm]{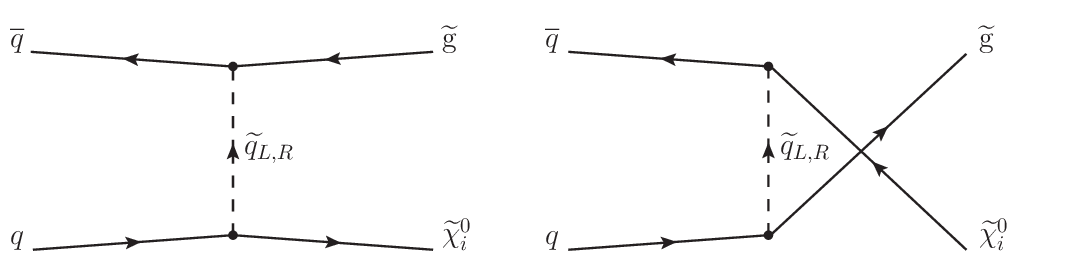}
\caption{ \label{fig:qqbarNsg} Feynman diagrams
of the subprocess $q\overline
q\rightarrow\widetilde\chi_{i}^{0}\widetilde{\text{g}}$ to leading
level.}
\end{figure}

 The production of neutralino-gluino originates from quark-antiquark initial states through
the tree-level Feynman diagrams shown in Fig.~\ref{fig:qqbarNsg} and can be expressed as
\begin{equation} \label{eq:qqng}
q^{k}(p_1)\overline
q^{l}(p_2)\rightarrow\widetilde\chi_{i}^{0}(k_1)\widetilde{\text{g}}^{m}(k_2),
\end{equation}
where $p_{1},p_{2},k_{1}$ and $k_{2}$ denote the four-momentum of the
quark, antiquark, and the two final-state neutralino and gluino,
respectively. Here, the color indices of the quark, antiquark, and gluino
are denoted by $k,l$ and $m$, respectively. The mass $m_j$ now
denotes the mass of the gluino in Eq.~\eqref{eq:cm} where the
kinematic is defined.

In this case, the production occurs by quark-antiquark scattering via
$t$-channel and  $u$-channel squark exchange in a
semi-strong reaction. The tree-level contributions to the amplitude
result from the two channels are
\begin{equation}  \label{eq:Tt_qqbarNsg}
\begin{split}
T_{\hat t}=&\frac{2g_s g T_{lj}^m}{(\hat
t-m_{\widetilde{q}_L}^2)c_W}\biggl[\overline{v}(p_2)P_R
u_{\widetilde{\text{g}}}^c(k_2)\biggr]\cdot\biggl[\overline{u}_{\widetilde
\chi_i^0}(k_1)C_{\widetilde
\chi_i^0 \widetilde q q}^{L} P_{L} u(p_1)\biggr]\\
&-\frac{2g_s g T_{lj}^m}{(\hat
t-m_{\widetilde{q}_R}^2)c_W}\biggl[\overline{v}(p_2)P_L
u_{\widetilde{\text{g}}}^c(k_2)\biggr]\cdot\biggl[\overline{u}_{\widetilde
\chi_i^0}(k_1)C_{\widetilde \chi_i^0 \widetilde q q}^{R} P_{R}
u(p_1)\biggr],
\end{split}
\end{equation}
\begin{equation}  \label{eq:Tu_qqbarNsg}
\begin{split}
T_{\hat u}=&-\frac{2g_s g T_{kj}^m}{(\hat
u-m_{\widetilde{q}_L}^2)c_W}\biggl[\overline{v}(p_2)C_{\widetilde
\chi_i^0 \widetilde q q}^{L*}P_R
u_{\widetilde{\chi}_i^0}^c(k_1)\biggr]\cdot\biggl[\overline{u}_{\widetilde{\text{g}}}(k_2) P_{L} u(p_1)\biggr]\\
&+\frac{2g_s g T_{kj}^m}{(\hat
u-m_{\widetilde{q}_R}^2)c_W}\biggl[\overline{v}(p_2)C_{\widetilde
\chi_i^0 \widetilde q q}^{R*}P_L
u_{\widetilde{\chi}_i^0}^c(k_1)\biggr]\cdot\biggl[\overline{u}_{\widetilde{\text{g}}}(k_2)
P_{R} u(p_1)\biggr],
\end{split}
\end{equation}
where the superscript $c$ denotes ``charge conjugate spinor''
defined by $\psi^c\equiv C\overline{\psi}^T$. In order to do the
spin sums, we use the spinor completeness relations given as $u=
C\overline{v}^T$ and $v= C\overline{u}^T$ for Majorana fermions
\cite{Haber}. The relevant couplings for this subprocess are given in Eq.
\eqref{eq:C_Nqsq}. After averaging over spins and colors in the
initial state, the analytic form of the partonic differential cross
section for this subprocess is obtained from these amplitudes by
using the following formula:
\begin{equation} \label{eq:dsigma_qqbarNsg}
\frac{d\hat \sigma }{d\hat t}(q\overline
q\rightarrow\widetilde\chi_{i}^{0}\widetilde{\text{g}})=\frac{1}{576\pi
\hat s^2}\left(M_{\hat t \hat t}+M_{\hat u \hat u}-2M_{\hat t \hat
u}\right),
\end{equation}
where
\begin{equation} \label{eq:Mtt_qqbarNsg}
M_{\hat t \hat
t}=\frac{16g_s^2g^2}{c_W^2}\biggl[\frac{|C_{\widetilde \chi_i^0
\widetilde q q}^{L}|^2}{(\hat
t-m_{\widetilde{q}_L}^2)^2}+\frac{|C_{\widetilde \chi_i^0 \widetilde
q q}^{R}|^2}{(\hat t-m_{\widetilde{q}_R}^2)^2}\biggr](m_{\widetilde
\chi_i^0}^2-\hat t)(m_{\widetilde{\text{g}}}^2-\hat t),
\end{equation}
\begin{equation} \label{eq:Muu_qqbarNsg}
M_{\hat u \hat
u}=\frac{16g_s^2g^2}{c_W^2}\biggl[\frac{|C_{\widetilde \chi_i^0
\widetilde q q}^{L}|^2}{(\hat
u-m_{\widetilde{q}_L}^2)^2}+\frac{|C_{\widetilde \chi_i^0 \widetilde
q q}^{R}|^2}{(\hat u-m_{\widetilde{q}_R}^2)^2}\biggr](m_{\widetilde
\chi_i^0}^2-\hat u)(m_{\widetilde{\text{g}}}^2-\hat u),
\end{equation}
\begin{equation} \label{eq:Mtu_qqbarNsg}
M_{\hat t \hat u}=\frac{16g_s^2g^2}{c_W^2}\biggl[\frac{C_{\widetilde
\chi_i^0 \widetilde q q}^{L} C_{\widetilde \chi_i^0 \widetilde q
q}^{L}}{(\hat t-m_{\widetilde{q}_L}^2)(\hat
u-m_{\widetilde{q}_L}^2)}+\frac{C_{\widetilde \chi_i^0 \widetilde q
q}^{R} C_{\widetilde \chi_i^0 \widetilde q q}^{R}}{(\hat
t-m_{\widetilde{q}_R}^2)(\hat
u-m_{\widetilde{q}_R}^2)}\biggr](m_{\widetilde \chi_i^0}
m_{\widetilde{\text{g}}} \hat s).
\end{equation}

\subsection{The subprocess $\boldsymbol{q \text{g} \rightarrow\widetilde\chi_{i}^{0}\widetilde{q}_{L,R}}$}
 The associated production of neutralino and squark, which can be produced via quark-gluon scattering, can be expressed through the following subprocess:
\begin{equation} \label{eq:qgNsq}
q^{k}(p_1)\text{g}^{l}(p_2)\rightarrow\widetilde\chi_{i}^{0}(k_1)\widetilde{q}_{L,R}^{m}(k_2),
\end{equation}
where $p_{1}$ and $p_{2}$ denote the four-momentum of the two initial-state quark and gluon,
and $k_{1}$ and $k_{2}$ denote the four-momentum of neutralino and squark in the final state, respectively. We denote
by $k, l$ and $m$ the color indices of the quark, gluon and squark,
respectively. In Eq. \eqref{eq:cm}, the mass $m_j$ now describes the
squark mass.

\begin{figure}[ht]
\includegraphics[width =12cm]{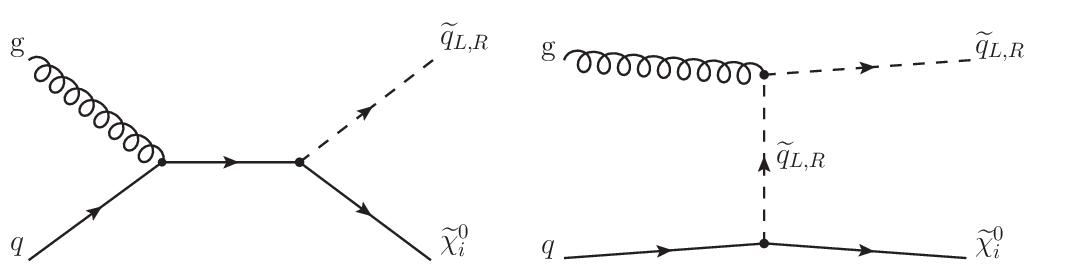}
\caption{ \label{fig:qgNsq} Feynman diagrams
of the subprocess $q \text{g}
\rightarrow\widetilde\chi_{i}^{0}\widetilde{q}_{L,R}$ to leading
level.}
\end{figure}

The tree-level Feynman diagrams of the subprocess are displayed in Fig.
\ref{fig:qgNsq}. This subprocess receives $s$-channel contribution
from exchange of quark, as well as $t$-channel contribution via
exchange of the left- and right-handed squark $\widetilde{q}_{L,R}$.
The leading-level contributions to the amplitude arising from the two
diagrams in Fig.~\ref{fig:qgNsq} are
\begin{equation}  \label{eq:Ts_qgNsq}
T_{\hat s}=-\frac{\sqrt 2 g_s g T_{kj}^l}{\hat s
c_W}\biggl[\overline{u}_{\widetilde{\chi}_i^0}(k_1)(C_{\widetilde
\chi_i^0 \widetilde q q}^{L} P_L+ C_{\widetilde \chi_i^0 \widetilde
q q}^{R} P_R)(\slashed{p}_1+ \slashed{p}_2)\slashed{\epsilon}_2
u(p_1)\biggr],
\end{equation}
\begin{equation}  \label{eq:Tt_qgNsq}
T_{\hat t}=-\frac{\sqrt 2 g_s g
T_{mj}^l}{c_W}\biggl[\overline{u}_{\widetilde{\chi}_i^0}(k_1)\biggl\{\frac{C_{\widetilde
\chi_i^0 \widetilde q q}^{L} P_L}{(\hat t-m_{\widetilde{q}_L}^2)}+
\frac{C_{\widetilde \chi_i^0 \widetilde q q}^{R} P_R}{(\hat
t-m_{\widetilde{q}_R}^2)}\biggr\}u(p_1)\biggr](2 \epsilon_2\cdot
k_2),
\end{equation}
where $\epsilon_2$ denotes the polarization vector of the initial gluon.
The relevant couplings are given in Eq.
\eqref{eq:C_Nqsq}. After averaging over spins and colors in the
initial state, the parton-level differential cross section for this
subprocess takes the form
\begin{equation} \label{eq:dsigma_qgNsq}
\frac{d\hat \sigma}{d\hat t}(q
\text{g}\rightarrow\widetilde\chi_{i}^{0}\widetilde{q}_{L,R})=\frac{1}{1536\pi
\hat s^2}\left(M_{\hat s \hat s}+M_{\hat t \hat t}+2M_{\hat s \hat
t}\right),
\end{equation}
where
\begin{equation} \label{eq:Mss_qgNsq}
M_{\hat s \hat s}=\frac{16g_s^2g^2}{\hat{s}^2
c_W^2}\biggl[|C_{\widetilde \chi_i^0 \widetilde q
q}^{L}|^2+|C_{\widetilde \chi_i^0 \widetilde q
q}^{R}|^2\biggr](m_{\widetilde \chi_i^0}^2-\hat u)\hat s,
\end{equation}
\begin{equation} \label{eq:Mtt_qgNsq}
M_{\hat t \hat
t}=\frac{32g_s^2g^2}{c_W^2}\biggl[\frac{|C_{\widetilde \chi_i^0
\widetilde q q}^{L}|^2 m_{\widetilde{q}_L}^2}{(\hat
t-m_{\widetilde{q}_L}^2)^2}+\frac{|C_{\widetilde \chi_i^0 \widetilde
q q}^{R}|^2 m_{\widetilde{q}_R}^2}{(\hat
t-m_{\widetilde{q}_R}^2)^2}\biggr](\hat t-m_{\widetilde
\chi_i^0}^2),
\end{equation}
\begin{equation} \label{eq:Mst_qgNsq}
M_{\hat s \hat t}=\frac{8g_s^2g^2}{c_W^2}\biggl[\frac{|C_{\widetilde
\chi_i^0 \widetilde q q}^{L}|^2 }{(\hat t-m_{\widetilde{q}_L}^2)\hat
s}+\frac{|C_{\widetilde \chi_i^0 \widetilde q q}^{R}|^2 }{(\hat
t-m_{\widetilde{q}_R}^2)\hat s}\biggr]\biggl[(m_{\widetilde
\chi_i^0}^2-\hat u)(m_{\widetilde \chi_i^0}^2-\hat t)-\hat
s(m_{\widetilde \chi_i^0}^2-\hat t)-\hat s m_{\widetilde
\chi_i^0}^2\biggr].
\end{equation}

\subsection{The subprocess $\boldsymbol{q\bar{q}^\prime\rightarrow\widetilde\chi_{i}^{0}
\widetilde\chi_{j}^{+}}$} 
\begin{figure}[h]
\includegraphics[width=12cm]{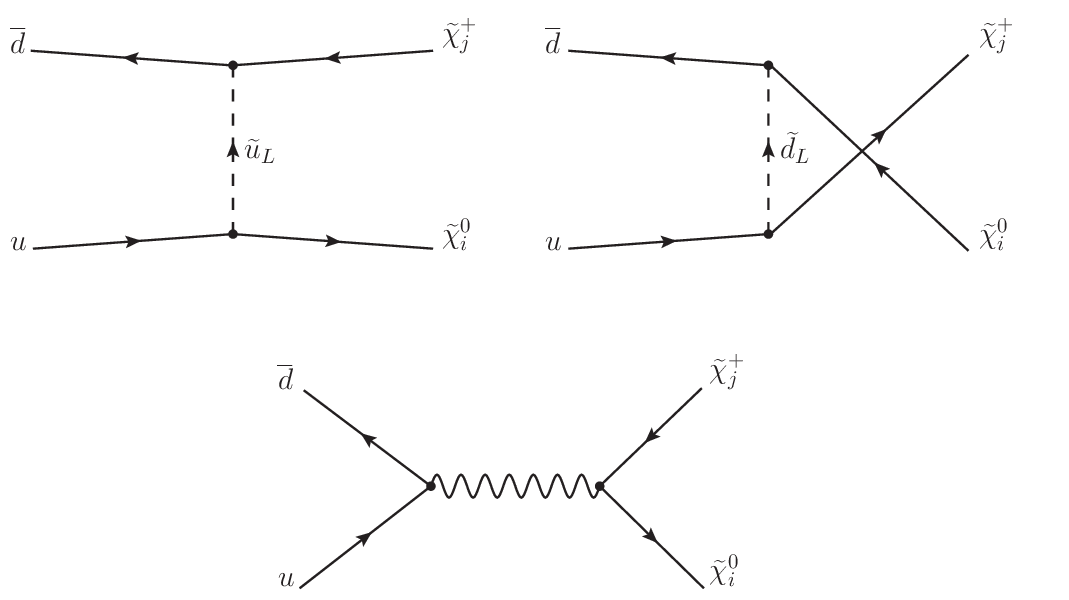}
\caption{ \label{fig:udbarNC} Feynman
diagrams of the subprocess $u\overline
d\rightarrow\widetilde\chi_{i}^{0}\widetilde\chi_{j}^{+}$ to leading
level.}
\end{figure}

The neutralino and chargino production,
which can dominantly be produced by annihilation of quarks and
antiquarks at hadron colliders as follows:
\begin{equation} \label{eq:qqbarNC}
q(p_1)\overline {q}^\prime
(p_2)\rightarrow\widetilde\chi_{i}^{0}(k_1)\widetilde\chi_{j}^{+}(k_2),
\end{equation}
where particle labels denote the corresponding four-momentum. The
kinematic is defined in Eq. \eqref{eq:cm}, with $m_i$ denoting the
neutralino mass and $m_j$ the chargino mass. The neutralino-chargino
production occurs via the Feynman diagrams shown in Fig.~\ref{fig:udbarNC}.
This subprocess proceeds at tree level via the vector boson $W^+$ exchange in the $s$-channel,
and via $t$- and $u$-channel exchange of the left squark $\widetilde{u}_L$ and $\widetilde{d}_L$.
The tree-level contributions to the amplitude arising from the three diagrams in
Fig.~\ref{fig:udbarNC} are
\begin{equation}  \label{eq:Ts_udbarNC}
\begin{split}
T_{\hat s}=-&\frac{g^2}{c_W^2}D_{W}(\hat
s)\biggl[\overline{u}_{i}(k_1) \gamma_{\mu}\left(O^{L}_{ij}P_L +
O^{R}_{ij}P_R\right){v}_{j} (k_2)\biggr]\\
&\cdot\biggl[\overline{v}(p_2)\gamma^{\mu}\left(L_{Wqq^{\prime}}P_L+R_{Wqq^{\prime}}P_R\right)u(p_1)\biggr],
\end{split}
\end{equation}
\begin{equation}  \label{eq:Tt_udbarNC}
\begin{split}
T_{\hat t}=&\frac{2g^2}{(\hat
t-m_{\widetilde{q}_L}^2)c_W^2}\biggl[\overline{u}_i(k_1)\left(C_{\widetilde
\chi_i^0 \widetilde q q}^{L*}P_L +C_{\widetilde
\chi_i^0 \widetilde q q}^{R*}P_R\right) u(p_1)\biggr]\\
&\cdot\biggl[\overline{v}(p_2)\gamma^{\mu}\left(C_{\widetilde \chi_j^+
\widetilde q q^{\prime}}^{L} P_{L}+C_{\widetilde \chi_j^+ \widetilde
q q^{\prime}}^{R}P_{R}\right)v_j(k_2)\biggr],
\end{split}
\end{equation}
\begin{equation}  \label{eq:Tu_udbarNC}
\begin{split}
T_{\hat u}=&\frac{2g^2}{(\hat
u-m_{{\widetilde{q}}^{\prime}_L}^2)c_W^2}\biggl[\overline{u}_j(k_2)\left(C_{\widetilde
\chi_j^+ {\widetilde{q}}^{\prime} q}^{L*} P_{L}+C_{\widetilde
\chi_j^+{\widetilde{q}}^{\prime} q}^{R*}P_{R}\right) u(p_1)\biggr]\\
&\cdot\biggl[\overline{v}(p_2)\gamma^{\mu}\left(C_{\widetilde \chi_i^0
{\widetilde{q}}^{\prime} q^{\prime}}^{L} P_{L}+C_{\widetilde
\chi_i^0 {\widetilde{q}}^{\prime}
q^{\prime}}^{R}P_{R}\right)v_i(k_1)\biggr].
\end{split}
\end{equation}

In order to obtain the cross section for this subprocess, one
would have to calculate the couplings of the neutralino-quark-squark,
chargino-quark-squark and neutralino-chargino-$W^+$ boson. We
summarize these couplings in Eqs.\eqref{eq:O_ij}-\eqref{eq:C_Cqsq}.
The analytic form of the partonic differential
cross section after spin and color averaging reads
\begin{equation} \label{eq:dsigma_udbarNC}
\frac{d\hat \sigma}{d\hat t}(q\overline
q^{\prime}\rightarrow\widetilde\chi_{i}^{0}\widetilde\chi_{j}^{+})=\frac{1}{192\pi
\hat s^2}\left(M_{\hat s \hat s}+M_{\hat t \hat t}+M_{\hat u \hat u}
-2M_{\hat s \hat t} +2M_{\hat s \hat u}-2M_{\hat t \hat u}\right),
\end{equation}
where
\begin{equation} \label{eq:Mss_udbarNC}
\begin{split}
M_{\hat s \hat s} =& \frac{4g^2|D_{W}(\hat s)|^2}{c_W^4}
\biggl\{\biggl[L_{Wqq^{\prime}}^2+R_{Wqq^{\prime}}^2\biggr]\biggl[O^{L}_{ij}O^{R*}_{ij}+
O^{L*}_{ij}O^{R}_{ij}\biggr]m_{\widetilde{\chi}_{i}^{0}}m_{\widetilde{\chi}_{j}^{+}}\hat
s\\
 &+\biggl[|O^{L}_{ij}|^2 L_{Wqq^{\prime}}^2+|O^{R}_{ij}|^2
 R_{Wqq^{\prime}}^2\biggr]
(m_{\widetilde{\chi}_{i}^{0}}^2 - \hat u)(m_{\widetilde{\chi}_{j}^{+}}^2 -\hat u) \\
& +\biggl[|O^{L}_{ij}|^2 R_{Wqq^{\prime}}^2+|O^{R}_{ij}|^2
L_{Wqq^{\prime}}^2\biggr] (m_{\widetilde{\chi}_{i}^{0}}^2 - \hat
t)(m_{\widetilde{\chi}_{j}^{+}}^2 -\hat t)\biggr\},
\end{split}
\end{equation}
\begin{equation} \label{eq:Mtt_udbarNC}
M_{\hat t \hat t}=\frac{4g^2}{(\hat t-m_{\widetilde{q}_L}^2)^2 c_W^4
} \biggl[|C_{\widetilde \chi_i^0 \widetilde q
q}^{L}|^2+|C_{\widetilde \chi_i^0 \widetilde q
q}^{R}|^2\biggr]\biggl[|C_{\widetilde \chi_j^+ \widetilde q
q^{\prime}}^{L}|^2+|C_{\widetilde \chi_j^+ \widetilde q
q^{\prime}}^{R}|^2\biggr](m_{\widetilde \chi_i^0}^2-\hat
t)(m_{\widetilde \chi_j^+}^2-\hat t),
\end{equation}
\begin{equation} \label{eq:Muu_udbarNC}
M_{\hat u \hat u}=\frac{4g^2}{(\hat
u-m_{\widetilde{q}^{\prime}_L}^2)^2 c_W^4 } \biggl[|C_{\widetilde
\chi_j^+ \widetilde{q}^{\prime} q}^{L}|^2+|C_{\widetilde \chi_j^+
\widetilde{q}^{\prime} q}^{R}|^2\biggr]\biggl[|C_{\widetilde
\chi_i^0 \widetilde{q}^{\prime} q^{\prime}}^{L}|^2+|C_{\widetilde
\chi_i^0 \widetilde{q}^{\prime}
q^{\prime}}^{R}|^2\biggr](m_{\widetilde \chi_i^0}^2-\hat
u)(m_{\widetilde \chi_j^+}^2-\hat u),
\end{equation}

$$
M_{\hat t \hat u}=\frac{4g^2}{(\hat t -
{m}_{\widetilde{q}_{L}}^2)(\hat
u-m_{\widetilde{q}^{\prime}_L}^2)c_W^4}
\biggl\{\frac{1}{2}\left[C_{\widetilde \chi_i^0 \widetilde q q}^{L*}
C_{\widetilde \chi_i^0 \widetilde{q}^{\prime} q^{\prime}}^{L*}
C_{\widetilde \chi_j^+ \widetilde{q}^{\prime} q}^{L} C_{\widetilde
\chi_j^+ \widetilde q q^{\prime}}^{L}+C_{\widetilde \chi_i^0
\widetilde q q}^{R*} C_{\widetilde \chi_i^0 \widetilde{q}^{\prime}
q^{\prime}}^{R*} C_{\widetilde \chi_j^+ \widetilde{q}^{\prime}
q}^{R} C_{\widetilde \chi_j^+ \widetilde q q^{\prime}}^{R} \right]
$$
\begin{equation}
\times\left[(m_{\widetilde\chi_{i}^{0}}^2-\hat
u)(m_{\widetilde{\chi}_{j}^{+}}^2-\hat u)+
(m_{\widetilde{\chi}_{i}^{0}}^2-\hat
t)(m_{\widetilde{\chi}_{j}^{+}}^2-\hat t)-\hat s(\hat s
-m_{\widetilde{\chi}_{i}^{0}}^2-m_{\widetilde{\chi}_{j}^{+}}^2)\right]
\end{equation}
$$
+m_{\widetilde{\chi}_{i}^{0}} m_{\widetilde{\chi}_{j}^{+}}\hat
s\left[C_{\widetilde \chi_i^0 \widetilde q q}^{L*} C_{\widetilde
\chi_i^0 \widetilde{q}^{\prime} q^{\prime}}^{R*} C_{\widetilde
\chi_j^+ \widetilde{q}^{\prime} q}^{L} C_{\widetilde \chi_j^+
\widetilde q q^{\prime}}^{R}+C_{\widetilde \chi_i^0 \widetilde q
q}^{R*} C_{\widetilde \chi_i^0 \widetilde{q}^{\prime}
q^{\prime}}^{L*} C_{\widetilde \chi_j^+ \widetilde{q}^{\prime}
q}^{R} C_{\widetilde \chi_j^+ \widetilde q q^{\prime}}^{L}
\right]\biggr\},
$$

$$
M_{\hat s \hat u}=\frac{-4g^4 (Re[D_{W}(\hat s)])}{(\hat
u-m_{\widetilde{q}^{\prime}_L}^2)c_W^4}\biggl\{\left[L_{Wqq^{\prime}}
O^{R}_{ij} C_{\widetilde \chi_i^0 \widetilde{q}^{\prime}
q^{\prime}}^{L*} C_{\widetilde \chi_j^+ \widetilde{q}^{\prime}
q}^{L}+ R_{Wqq^{\prime}} O^{L}_{ij} C_{\widetilde \chi_i^0
\widetilde{q}^{\prime} q^{\prime}}^{R*} C_{\widetilde \chi_j^+
\widetilde{q}^{\prime} q}^{R} \right]\\
$$
\begin{equation}
\times(m_{\widetilde{\chi}_{i}^{0}}^2-\hat
u)(m_{\widetilde{\chi}_{j}^{+}}^2 -\hat u)+ m_{\widetilde{\chi}_{i}^{0}} m_{\widetilde{\chi}_{j}^{+}} \hat s \\
\end{equation}
$$
\times\left[L_{Wqq^{\prime}} O^{L}_{ij} C_{\widetilde \chi_i^0
\widetilde{q}^{\prime} q^{\prime}}^{L*} C_{\widetilde \chi_j^+
\widetilde{q}^{\prime} q}^{L}+ R_{Wqq^{\prime}} O^{R}_{ij}
C_{\widetilde \chi_i^0 \widetilde{q}^{\prime} q^{\prime}}^{R*}
C_{\widetilde \chi_j^+ \widetilde{q}^{\prime} q}^{R} \right]
\biggr\},
$$

$$
M_{\hat s \hat t}=\frac{-4g^4 (Re[D_{W}(\hat s)])}{(\hat
t-m_{\widetilde{q}_L}^2)c_W^4}\biggl\{\left[L_{Wqq^{\prime}}
O^{R}_{ij} C_{\widetilde \chi_i^0 \widetilde{q} q}^{L} C_{\widetilde
\chi_j^+ \widetilde{q} q^{\prime}}^{L*}+ R_{Wqq^{\prime}} O^{L}_{ij}
C_{\widetilde \chi_i^0 \widetilde{q} q}^{R} C_{\widetilde
\chi_j^+ \widetilde{q} q^{\prime}}^{R*} \right]\\
$$
\begin{equation}
\times(m_{\widetilde{\chi}_{i}^{0}}^2-\hat
t)(m_{\widetilde{\chi}_{j}^{+}}^2 -\hat t)+ m_{\widetilde{\chi}_{i}^{0}} m_{\widetilde{\chi}_{j}^{+}} \hat s \\
\end{equation}
$$
\times\left[L_{Wqq^{\prime}} O^{L}_{ij} C_{\widetilde \chi_i^0
\widetilde{q} q}^{L} C_{\widetilde \chi_j^+ \widetilde{q}
q^{\prime}}^{L*}+ R_{Wqq^{\prime}} O^{R}_{ij} C_{\widetilde \chi_i^0
\widetilde{q} q}^{R} C_{\widetilde \chi_j^+ \widetilde{q}
q^{\prime}}^{R*} \right] \biggr\}.
$$
In the above relations, the following abbreviation has been used $
D_{W}(\hat s)=\frac{1}{\hat s - m_{W}^2+im_{W} \Gamma_{W}} $, which
is the $W$-boson propagator denominator. We get $m_{W}$ = 80.385
GeV and the width of this boson is $\Gamma_{W}=2.085$ GeV for
calculations.

\subsection{The subprocess $\boldsymbol{\text{g}\text{g}\rightarrow\widetilde\chi_{i}^{0}\widetilde{\text{g}}}$}
\begin{figure} [ht]
\centering \epsfig{figure=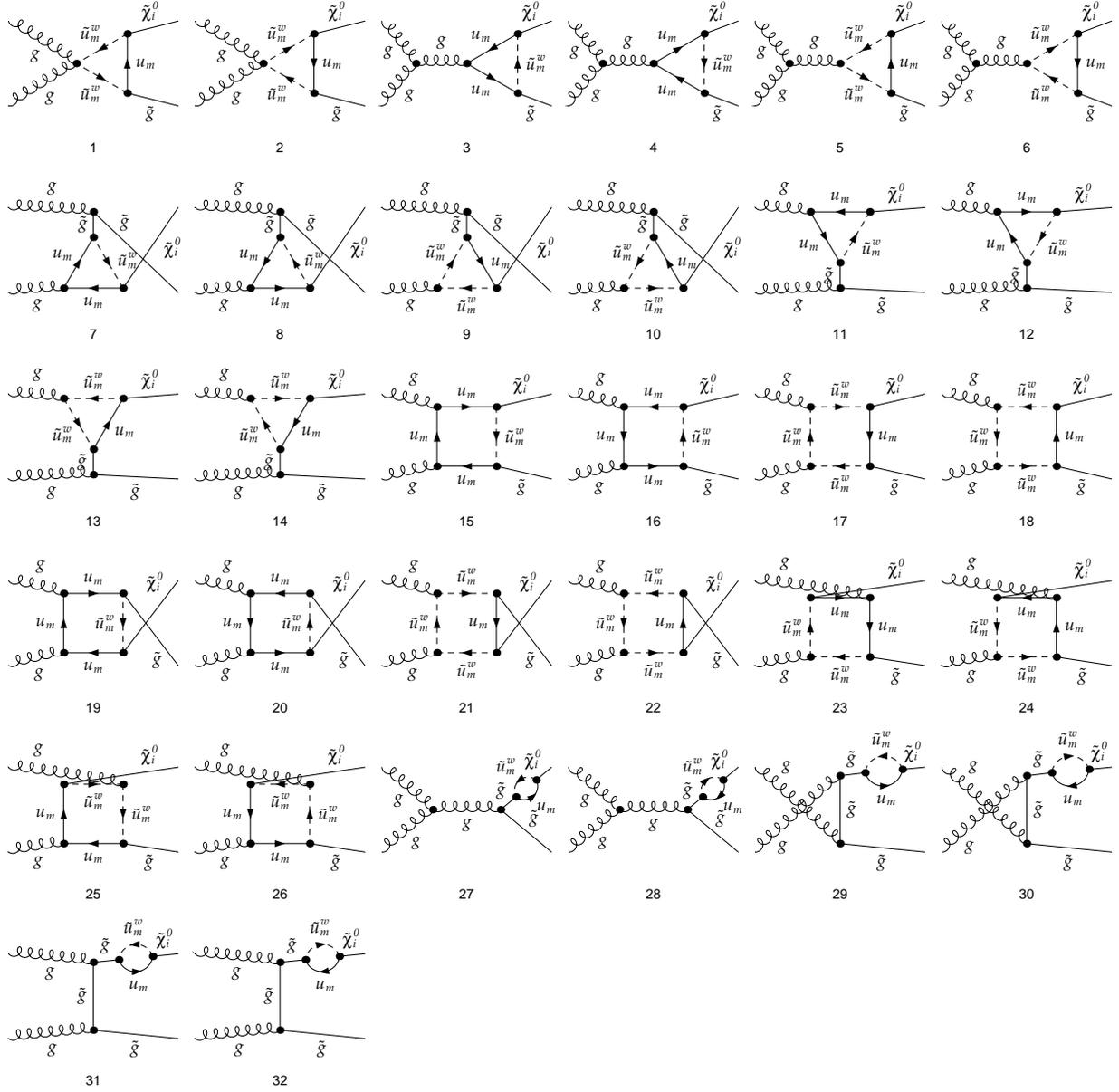,width=1\textwidth} \caption{Feynman
diagrams of the subprocess
$\text{g}\text{g}\rightarrow\widetilde\chi_{i}^{0}\widetilde{\text{g}}$
to one-loop level. Also, this subprocess contains diagrams which
are obtained by the replacements $u_m \rightarrow d_m$ and
$\widetilde{u}_m^w\rightarrow\widetilde{d}_m^w$ in the above
diagrams. Here, \textit{m} and \textit{w} indices denote the
generation of (s)quark and the mass eigenstate of squark,
respectively.} \label{fig:ggNsg}
\end{figure}

 The associated production of neutralino and gluino can be produced via the
collision of gluon-gluon as follows:
\begin{equation} \label{eq:ggNsg}
\text{g}^{k}(p_1)\text{g}^{l}(p_2)\rightarrow\widetilde\chi_{i}^{0}(k_1)\widetilde{\text{g}}^{m}(k_2),
\end{equation}
where $p_{1}$ and $p_{2}$ denote the four-momentum of the initial gluons,
and $k_{1}$ and $k_{2}$ represent the four-momentum of the two
final-state neutralino and gluino, respectively. We denote by $k, l$
and $m$ the color indices of gluons and gluino, respectively. This
subprocess first emerges at the one-loop level. We have performed the
numerical evaluation for the subprocess $\text{g} \text{g}\to
\widetilde{\chi}_{i}^{0}\widetilde{\text{g}}$ at one-loop using the
Mathematica packages \verb"FeynArts" \cite{Feynarts} to calculate
corresponding amplitudes, \verb"FormCalc" \cite{Hahn,Hahn2} to
produce a complete Fortran code containing the squared matrix
elements, and \verb"LoopTools" \cite{loop} to perform the evaluation
of the necessary loop integrals. Also, the Feynman diagrams depicted in
Fig.~\ref{fig:ggNsg} have been generated by using \verb"FeynArts".
In general, the one-loop corrections to subprocess
$\text{g}\text{g}\rightarrow\widetilde\chi_{i}^{0}\widetilde{\text{g}}$
could be classified as vertex contributions and box contributions.
The calculations of this subprocess have been carried out in the 't
Hooft-Feynman gauge in which the gluon polarization sum is
$\sum_\lambda
\epsilon_\mu^*(k,\lambda)\epsilon_\nu(k,\lambda)=-g_{\mu\nu}$.
For regularization of the ultraviolet divergences, we have used
the constrained differential renormalization (CDR) \cite{CDR}, which
has been shown to be equivalent to regularization by dimensional
reduction \cite{DR,DR2} at the one-loop level. Therefore, a
supersymmetry-preserving regularization scheme is ensured via the
implementation given in Ref. \cite{DR3}. We do not display the
analytical results of this process due to the fact that these are too long to
be included here.

\newpage

\section{Numerical analysis and discussions}\label{results}

We now present numerical predictions for the cross sections of
the single neutralino production in $pp$ collisions at the LHC energies.
We investigate the direct production of a single neutralino
$\widetilde{\chi}_{i}^{0}$ for first-generation quarks at hadron
colliders focusing on the $\widetilde{\chi}_{1}^{0}$ is likely to
be the LSP and $\widetilde{\chi}_{2}^{0}$. The relevant subprocesses
are $q \bar{q} \to \widetilde {\chi}_{i}^{0}\widetilde{\text{g}}$,
$q \text{g}\to \widetilde {\chi}_{i}^{0}\widetilde{q}_{L,R}$ and $q
\bar{q}^\prime \to \widetilde
{\chi}_{i}^{0}\widetilde{\chi}_{j}^{\pm}$ at tree-level, while
$\text{g} \text{g}\to \widetilde{\chi}_{i}^{0}\widetilde{\text{g}}$
at one-loop level, which could lead to the first detection of the supersymmetric particles at the LHC.
In the numerical calculations, we just limit
the values of the mass parameters $M_1$, $M_2$ and $\mu$ to be real, positive and below
1 TeV, and get $\tan\beta$= 45, $m_{\widetilde{u}_{R}}$= 799.2 GeV,
$m_{\widetilde{u}_{L}}$= 798.2 GeV, $m_{\widetilde{d}_{R}}$= 802.3
GeV, $m_{\widetilde{d}_{L}} $= 800.3 GeV and $m_{\widetilde{g}} $=
1400 GeV. For the other parameters, we use the values given by
the Particle Data Group, such as $m_Z$= 91.1876 GeV, $m_W$= 80.399
GeV \cite{PDG}.  By using Eqs.~\eqref{eq:2M2} and \eqref{eq:2mu2}
with two chargino masses, one could have three choices of parameter
sets for the gaugino/Higgsino mass parameters $M_2$ and $\mu$ in three different cases, which are the
gaugino-like, the higgsino-like and the mixture-case, respectively.
We fix masses of the charginos as $m_{\widetilde{\chi}_{1}^{\pm}}=
 168.51$ GeV and $m_{\widetilde{\chi}_{2}^{\pm}}= 295.01$ GeV for
gaugino and higgsino-like scenarios, and
$m_{\widetilde{\chi}_{1}^{\pm}}= 173.66$ GeV and
$m_{\widetilde{\chi}_{2}^{\pm}}= 289.86$ GeV for mixture-case. For
each scenario, neutralino masses are calculated by inserting the
values of $M_2$ and $\mu$ into Eq.~\eqref{eq:mN}.~\cref{table1} shows the gaugino/Higgsino and neutralino masses.
\begin{table}[htp]
\caption{The gaugino/Higgsino mass parameters and neutralino masses
for each scenario.}\label{table1}
\begin{ruledtabular}
\begin{tabular}{lR[.][.]{3}{2}R[.][.]{3}{2}R[.][.]{3}{2}R[.][.]{3}{2}R[.][.]{3}{2}R[.][.]{3}{2}R[.][.]{3}{2}}
 ~~~~[in GeV]&\multicolumn{1}{c}{$M_{2}$}&\multicolumn{1}{c}{$\mu$}&\multicolumn{1}{c}{$M_{1}$}&\multicolumn{1}{c}{$m_{\widetilde{\chi}_{1}^{0}}$}& \multicolumn{1}{c}{$m_{\widetilde{\chi}_{2}^{0}}$}&\multicolumn{1}{c}{$m_{\widetilde{\chi}_{3}^{0}}$}&\multicolumn{1}{c}{$m_{\widetilde{\chi}_{4}^{0}}$}\\
\hline
Higgsino like&250.00&200.00&119.33&109.59&174.50&209.65&294.88\\
Gaugino like& 200&250.00& 95.46 &91.50&169.50&259.40&293.85\\
Mixture case&225.00&225.00&107.39 &101.42&176.13&234.52&289.37\\
CMSSM 40.2.2~~& 391.24&698.59& 210.84 &208.23& 397.26&702.97&711.31\\
\end{tabular}
\end{ruledtabular}
\end{table}

For comparison, we have also worked out the cross sections in the
CMSSM 40.2.2 benchmark point \cite{CMSSM4022} in the framework of
the CMSSM \cite{CMSSM,CMSSM2,CMSSM3} with five
input parameters, namely, $m_0=$ 600 GeV, $m_{1/2}=$ 500 GeV,
$A_0=-$ 500 GeV, $\tan\beta=$ 40 and $\mu>0$, where  the parameters
$m_{0}$ and $m_{1/2}$ are the universal scalar and gaugino mass
parameters, $A_0$ is the universal trilinear soft SUSY breaking
parameter, $\tan\beta$ is the ratio of the vacuum expectation values
of the two Higgs doublets and sign($\mu$) is the sign of the Higgs
mixing parameter. The universal parameters $m_{0}$, $m_{1/2}$ and
$A_0$ are thought to appear by means of some gravity-mediated
mechanism and are defined at the grand unified theories scale, whereas $\tan\beta$ and
sign of the Higgs mixing parameter sign($\mu$) are defined at the
electroweak scale. All the masses and
couplings of the model from these five parameters are obtained by the evolution from the grand unified theories scale down to the electroweak scale \cite{Drees}. In this case,
we have computed the SUSY particle spectrum by using
\verb"SoftSusy-3.3.4" package \cite{softsusy}. For the CMSSM 40.2.2
benchmark point, the gaugino masses $M_2$ and $M_1$, the Higgsino
mass $\mu$, and neutralino masses are given in~\cref{table1}, and the
other parameters are obtained as
$m_{\widetilde{\chi}_{1}^{+}}=397.33$ GeV,
 $m_{\widetilde{\chi}_{2}^{+}}=711.85$
GeV, $m_{\widetilde{u}_{L}}=1199.95$ GeV,
$m_{\widetilde{d}_{L}}=1202.41$ GeV, $m_{\widetilde{u}_{R}}=1167.94$
GeV, $m_{\widetilde{d}_{R}}=1165.21$ GeV, and
$m_{\widetilde{\text{g}}}=1170.38$ GeV.

\begin{figure}[hpt]
    \begin{center}
\includegraphics[scale=0.36]{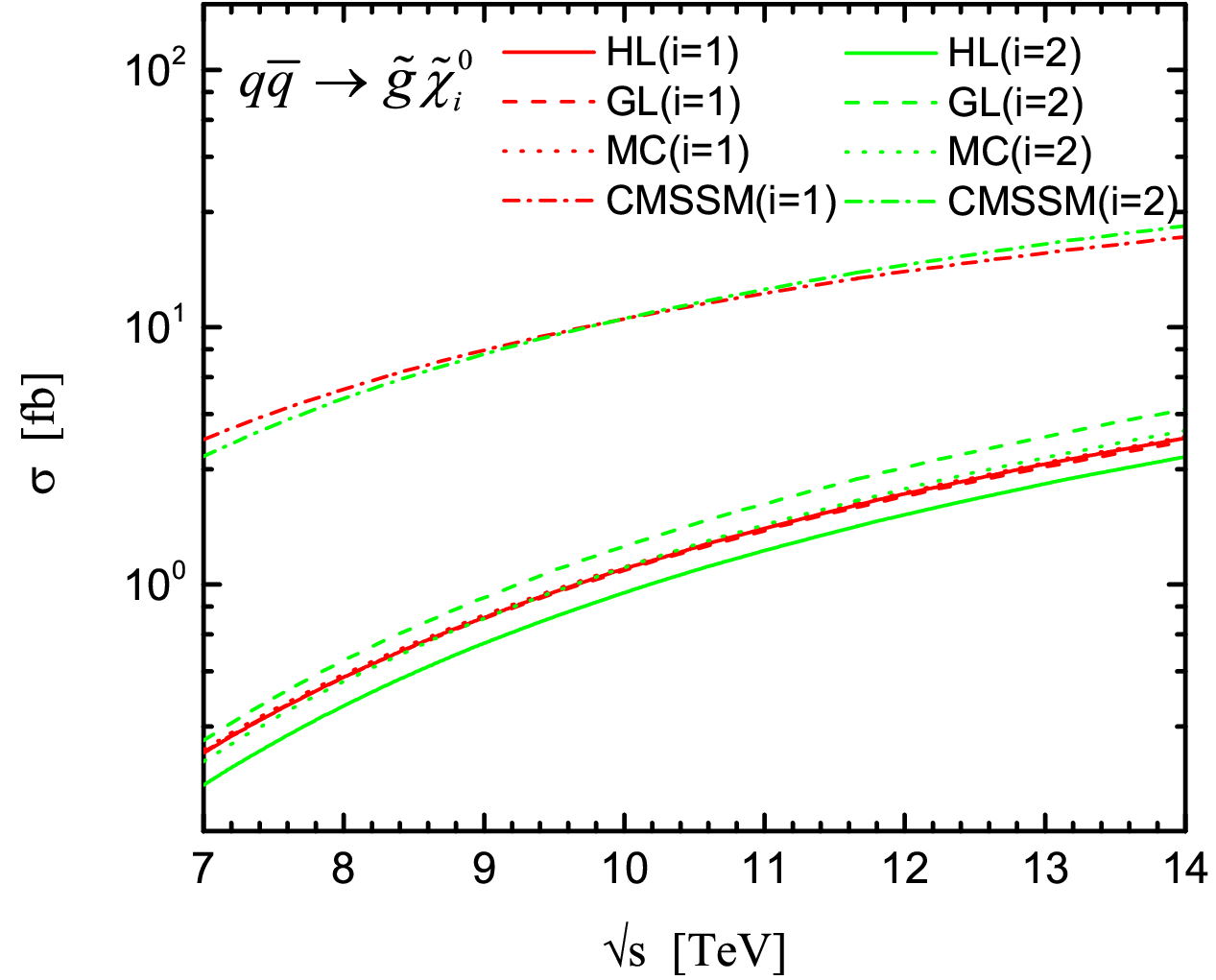}
     \end{center}
\caption{(color online). Total cross sections of the process
$pp\to\widetilde{\text{g}}\widetilde{\chi}_{i}^0$ (i=1,2) versus the
center-of-mass energy of $pp$ collider $\sqrt{s}$.} \label{Fig5}
\end{figure}

\begin{figure}[hpt]
    \begin{center}
\includegraphics[scale=0.36]{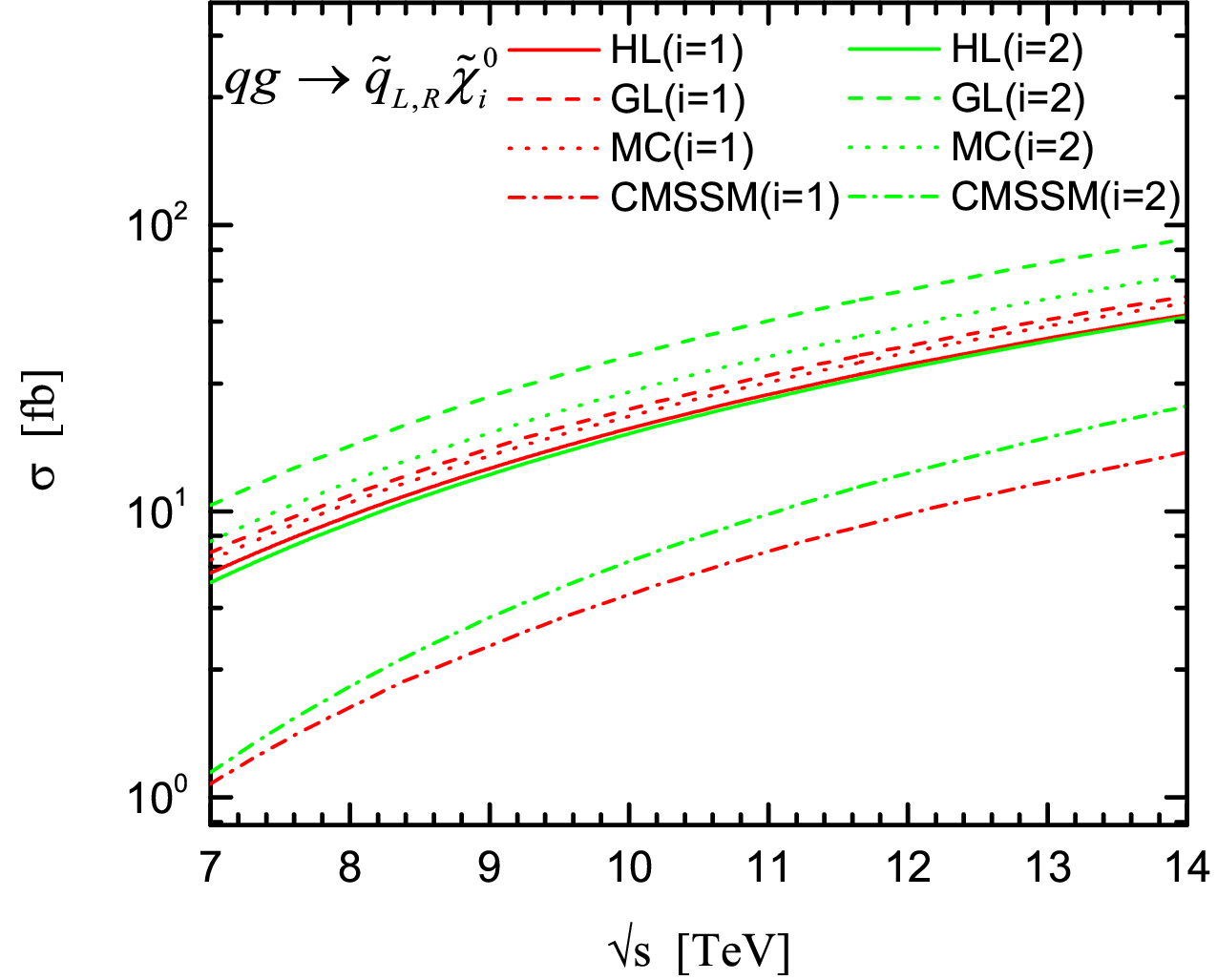}
     \end{center}
\caption{(color online). Total cross sections for the process
$pp\to\widetilde{q}_{L,R}\widetilde{\chi}_{i}^0$ (i=1,2) versus the
center-of-mass energy of $pp$ collider $\sqrt{s}$.} \label{Fig6}
\end{figure}

\begin{figure}[hpt]
    \begin{center}
\includegraphics[scale=0.36]{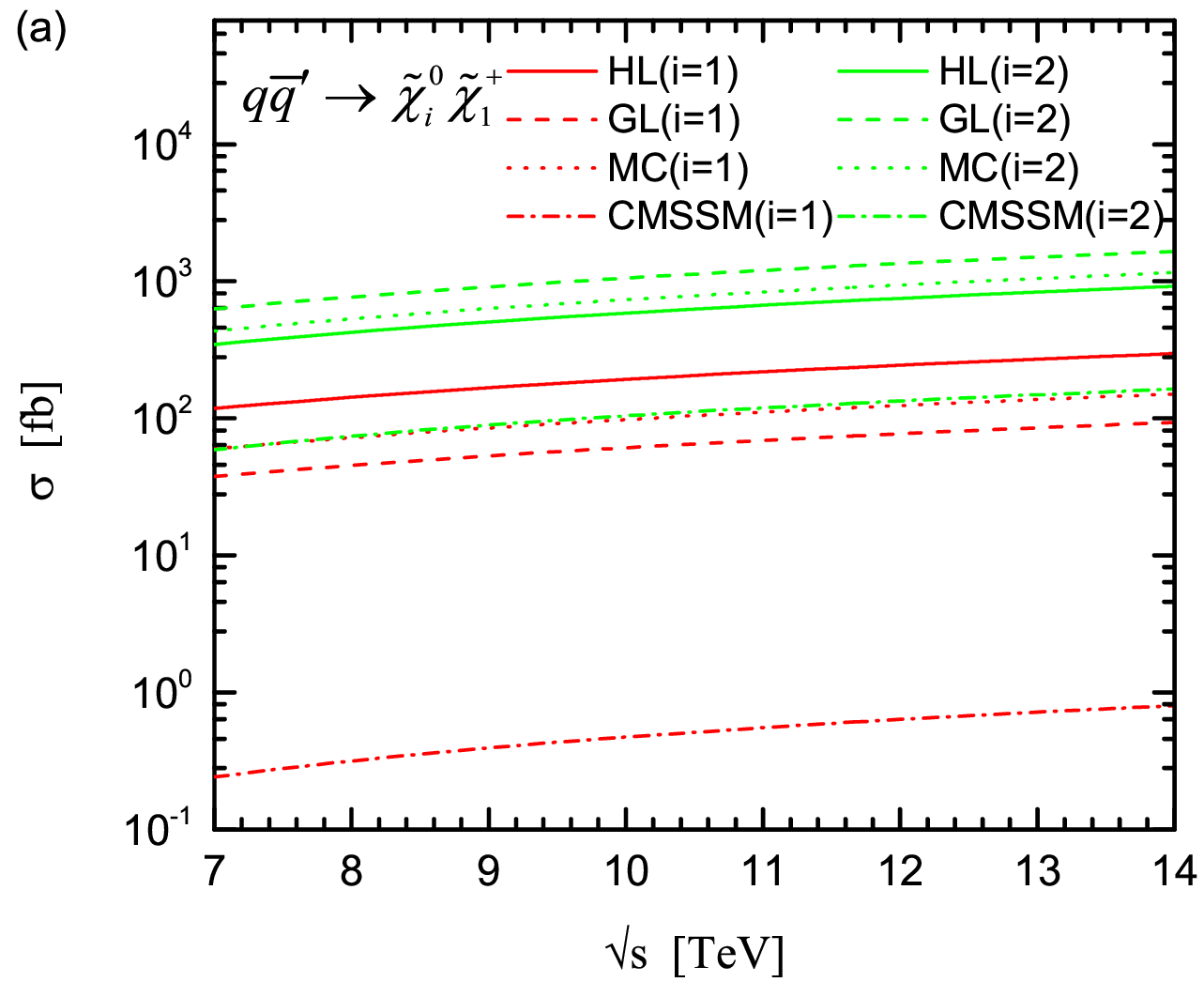}
\includegraphics[scale=0.36]{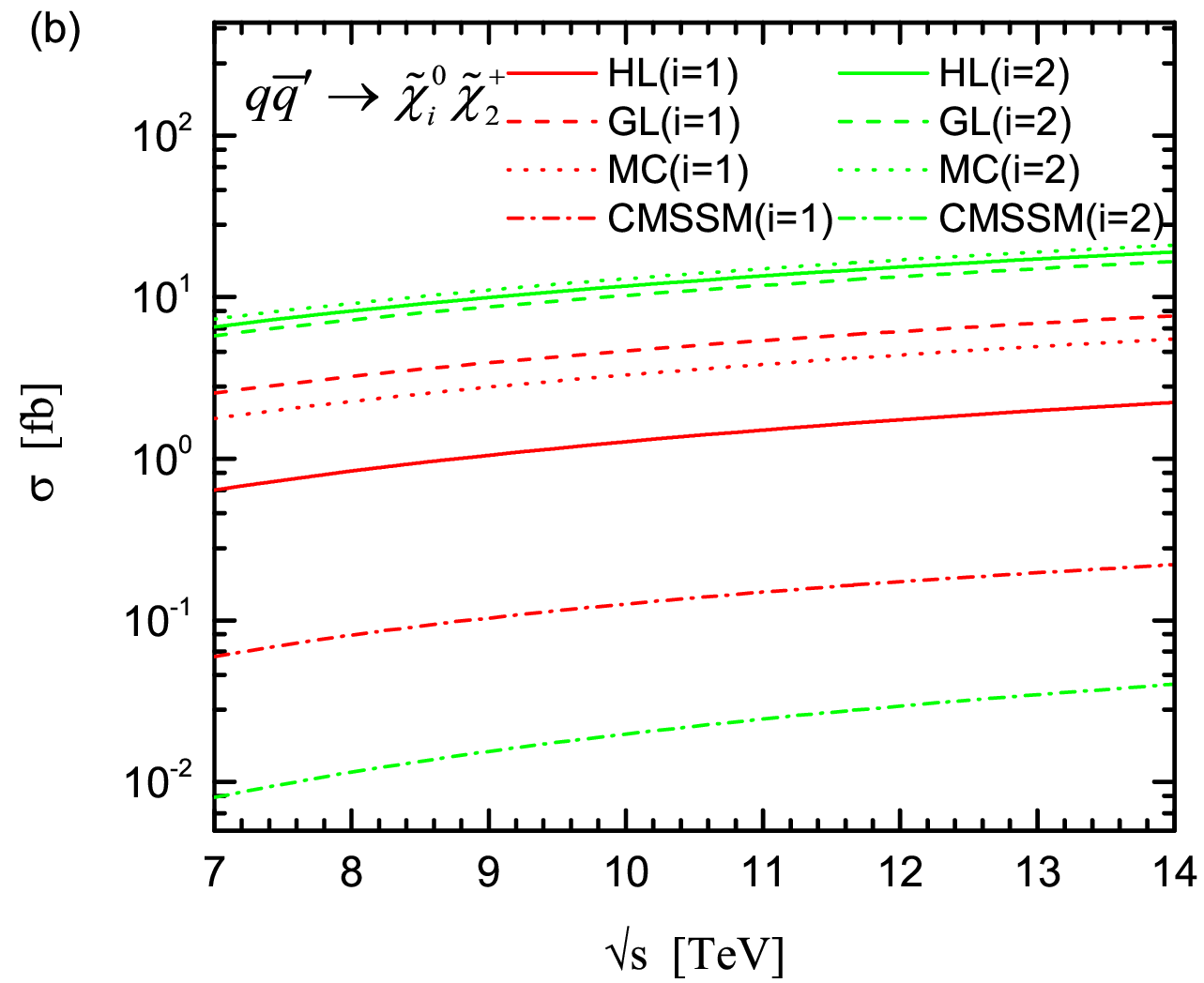}
      \end{center}
\caption{(color online). Total cross sections of the processes $pp
\to\widetilde{\chi}_{i}^0\widetilde{\chi}_{1}^+$ (left) and
$\widetilde{\chi}_{i}^0\widetilde{\chi}_{2}^+$ (right) (i=1,2)
versus the center-of-mass energy of $pp$ collider $\sqrt{s}$.}
\label{Fig7}
\end{figure}

\begin{figure}[hpt]
    \begin{center}
\includegraphics[scale=0.36]{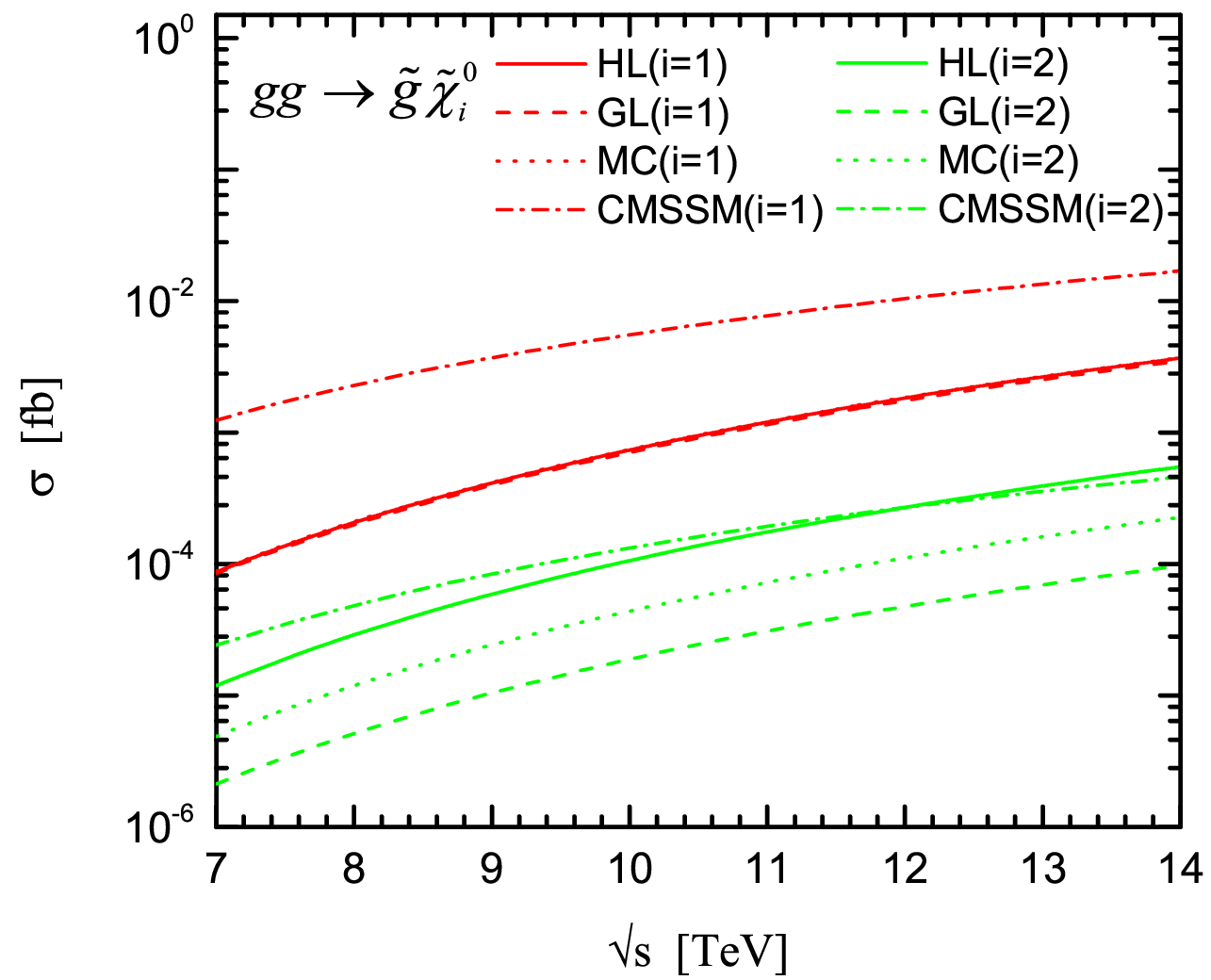}
     \end{center}
\caption{(color online). Total cross sections of the process
$pp(\text{g}\text{g})\to\widetilde{\text{g}}\widetilde{\chi}_{i}^0$
(i=1,2) versus the center-of-mass energy of $pp$ collider
$\sqrt{s}$.} \label{Fig8}
\end{figure}

\begin{figure}[hpt]
    \begin{center}
\includegraphics[scale=0.36]{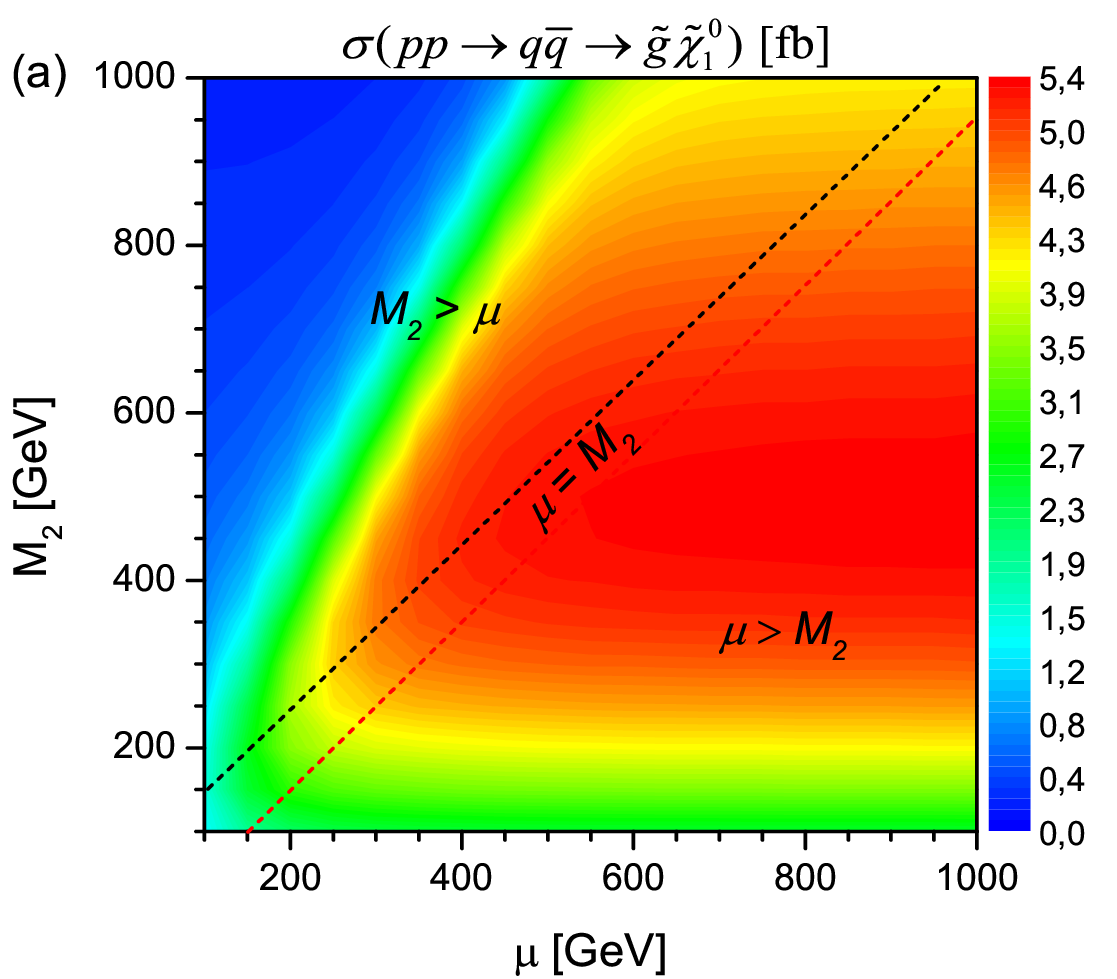}
\includegraphics[scale=0.36]{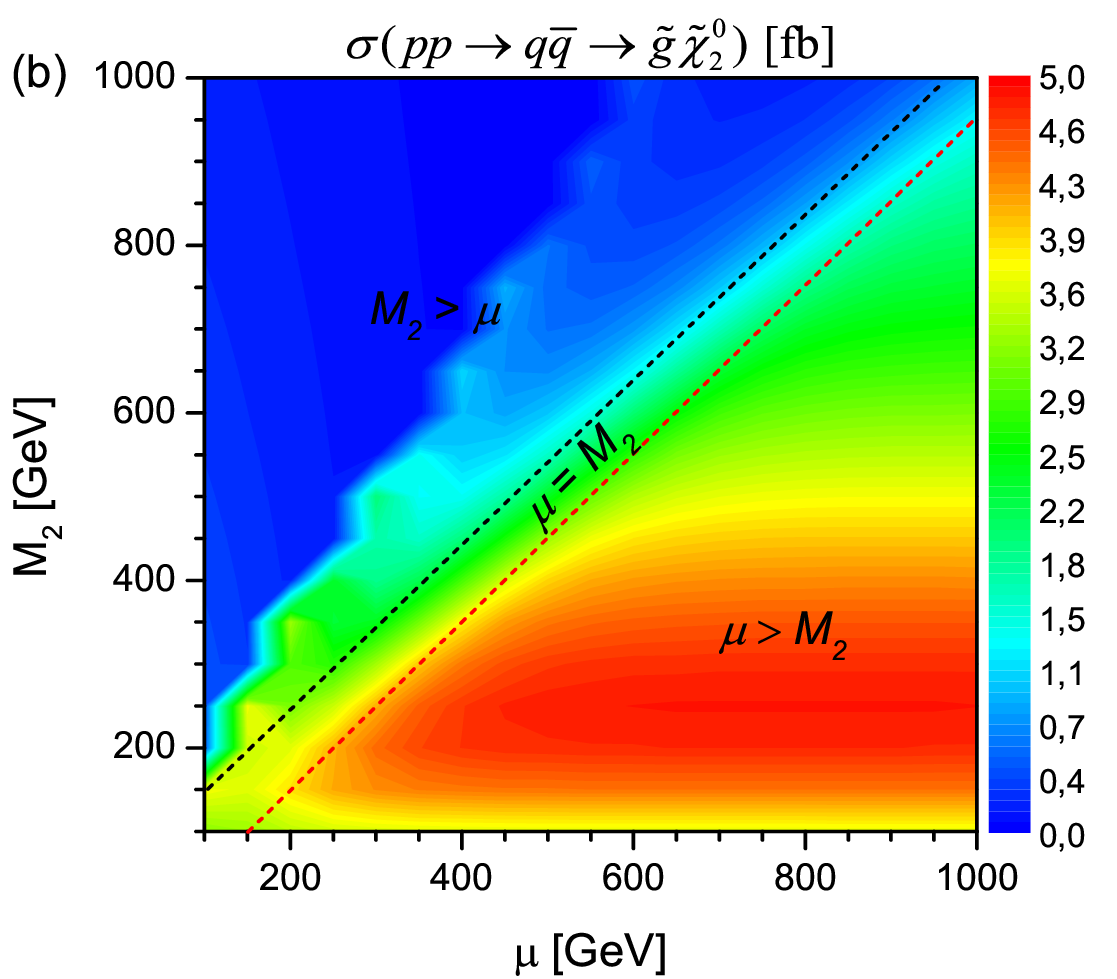}
     \end{center}
\caption{(color online). Contour plots of the total cross sections
of the process $pp\to\widetilde{\text{g}}\widetilde{\chi}_{i}^0$
(i=1,2) in the $M_2-\mu$ plane for $\sqrt{s}=8$ TeV. We choose
$\tan\beta=45$ and fix $M_1=\frac{5}{3}M_2 \tan^2\theta_W$.}
\label{Fig9}
\end{figure}

\begin{figure}[hpt]
    \begin{center}
\includegraphics[scale=0.36]{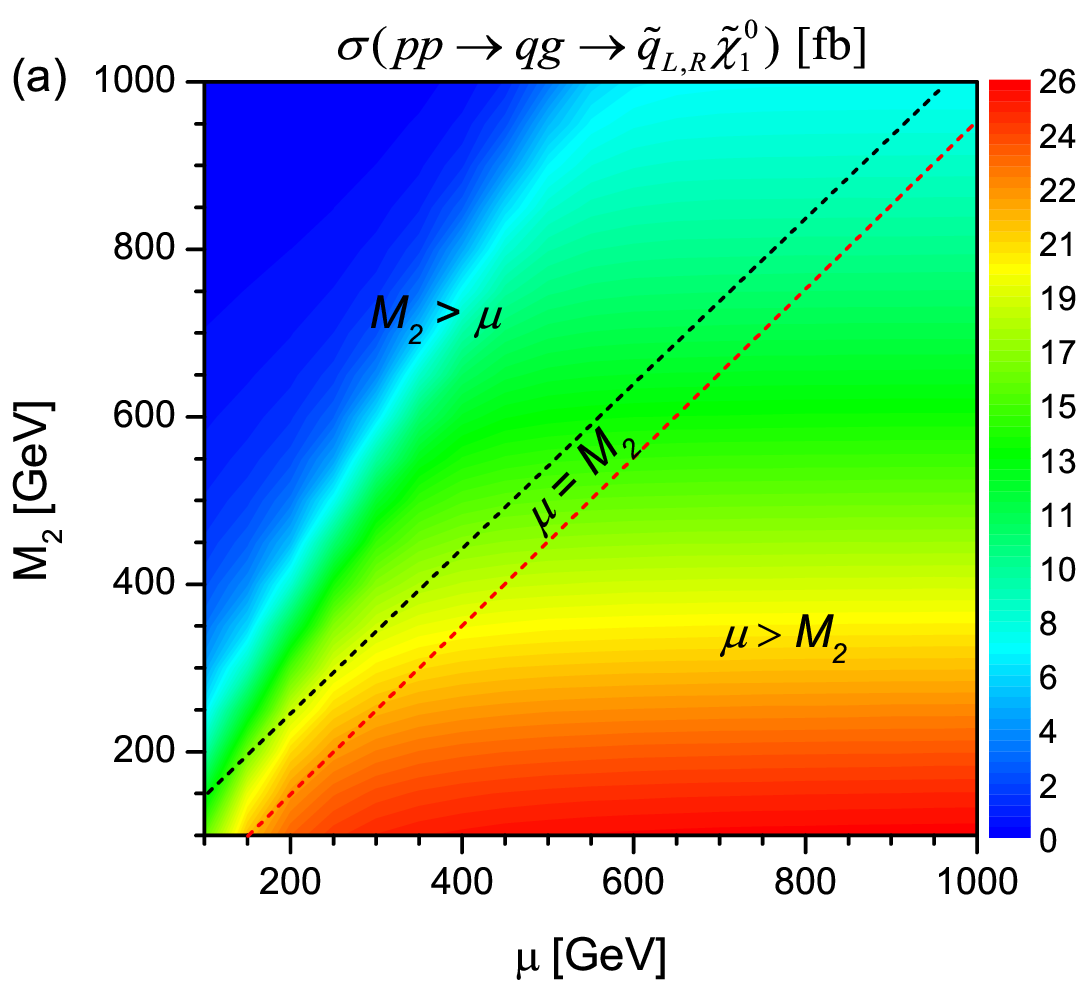}
\includegraphics[scale=0.36]{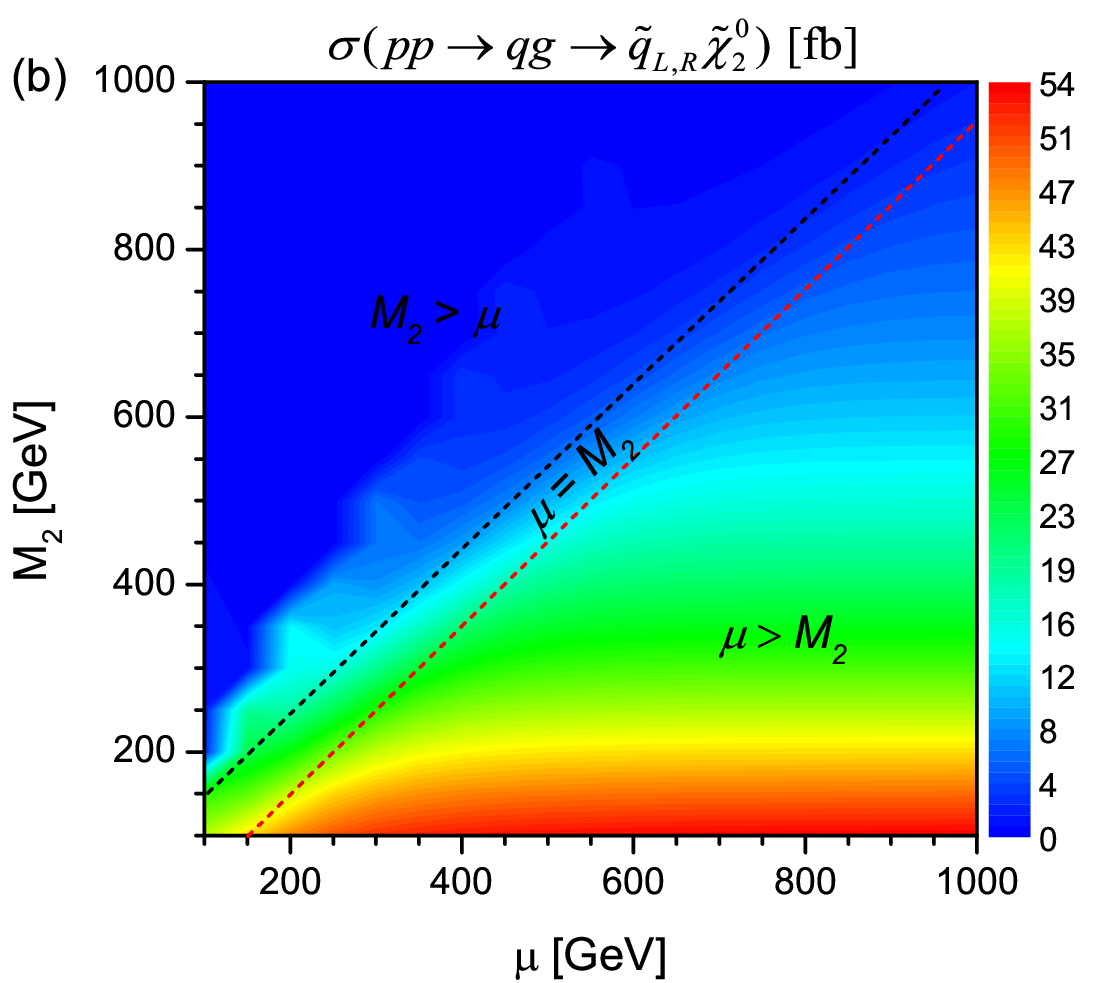}
     \end{center}
\caption{(color online). Contour plots of the total cross sections
of the process $pp\to\widetilde{q}_{L,R}\widetilde{\chi}_{i}^0$
(i=1,2) in the $M_2-\mu$ plane for $\sqrt{s}=8$ TeV. We choose
$\tan\beta=45$ and fix $M_1=\frac{5}{3}M_2 \tan^2\theta_W$.}
\label{Fig10}
\end{figure}

\begin{figure}[hpt]
    \begin{center}
\includegraphics[scale=0.35]{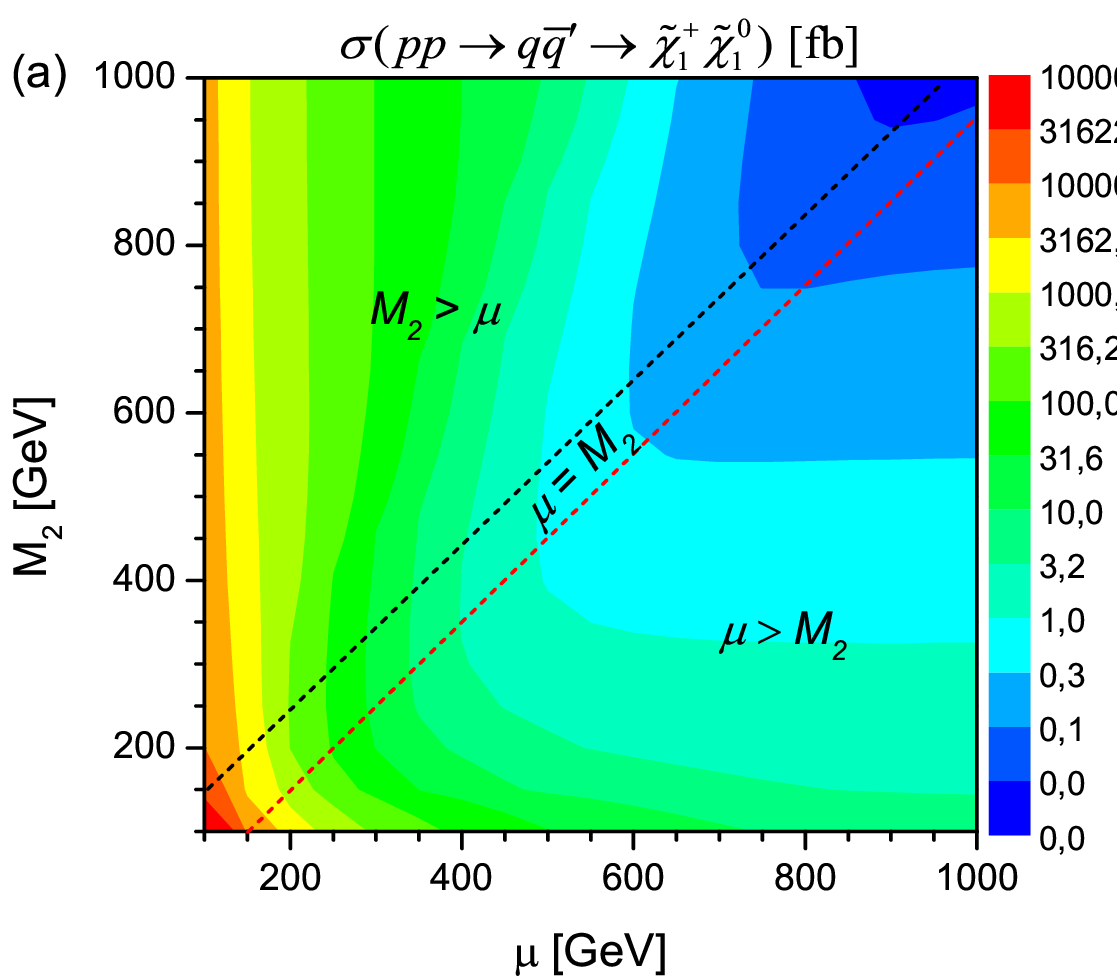}
\includegraphics[scale=0.35]{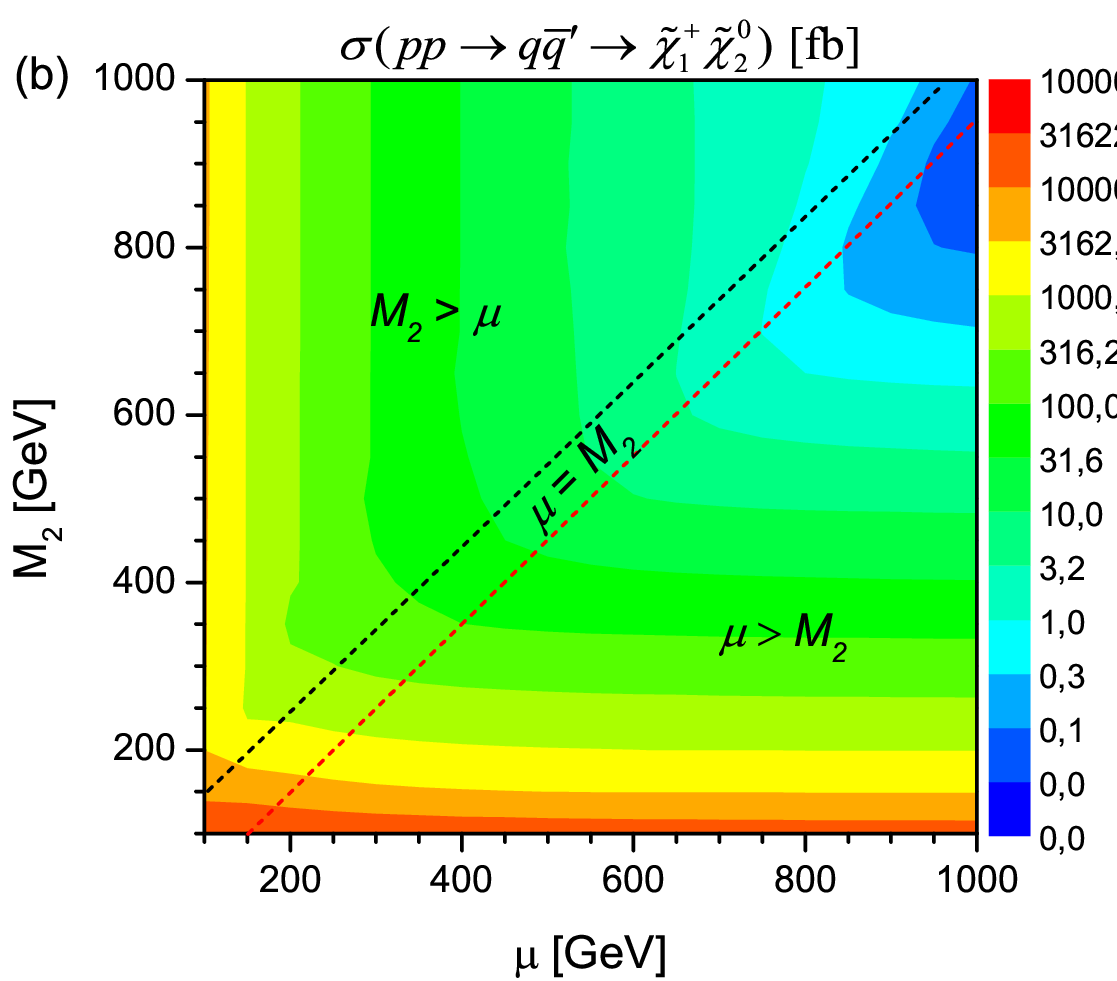}
\includegraphics[scale=0.35]{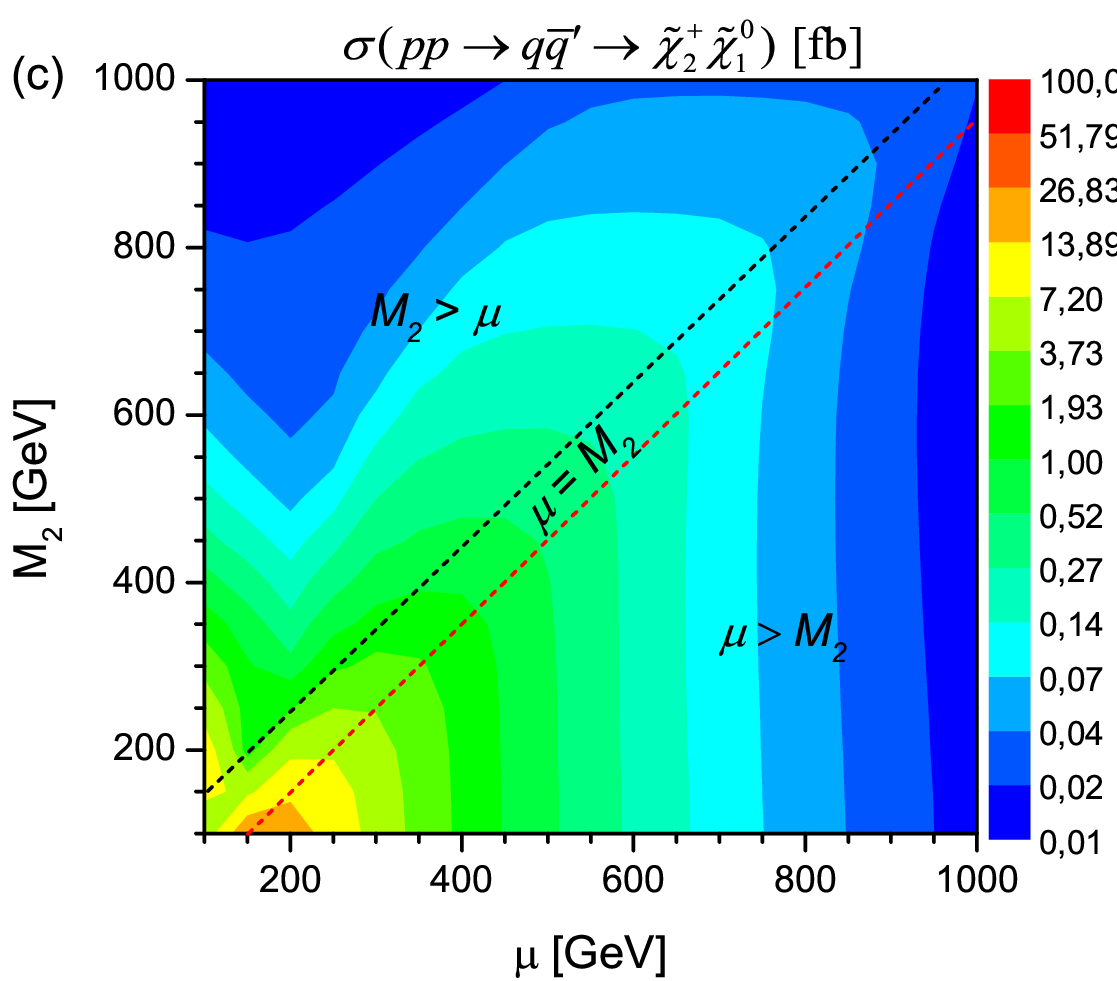}
\includegraphics[scale=0.35]{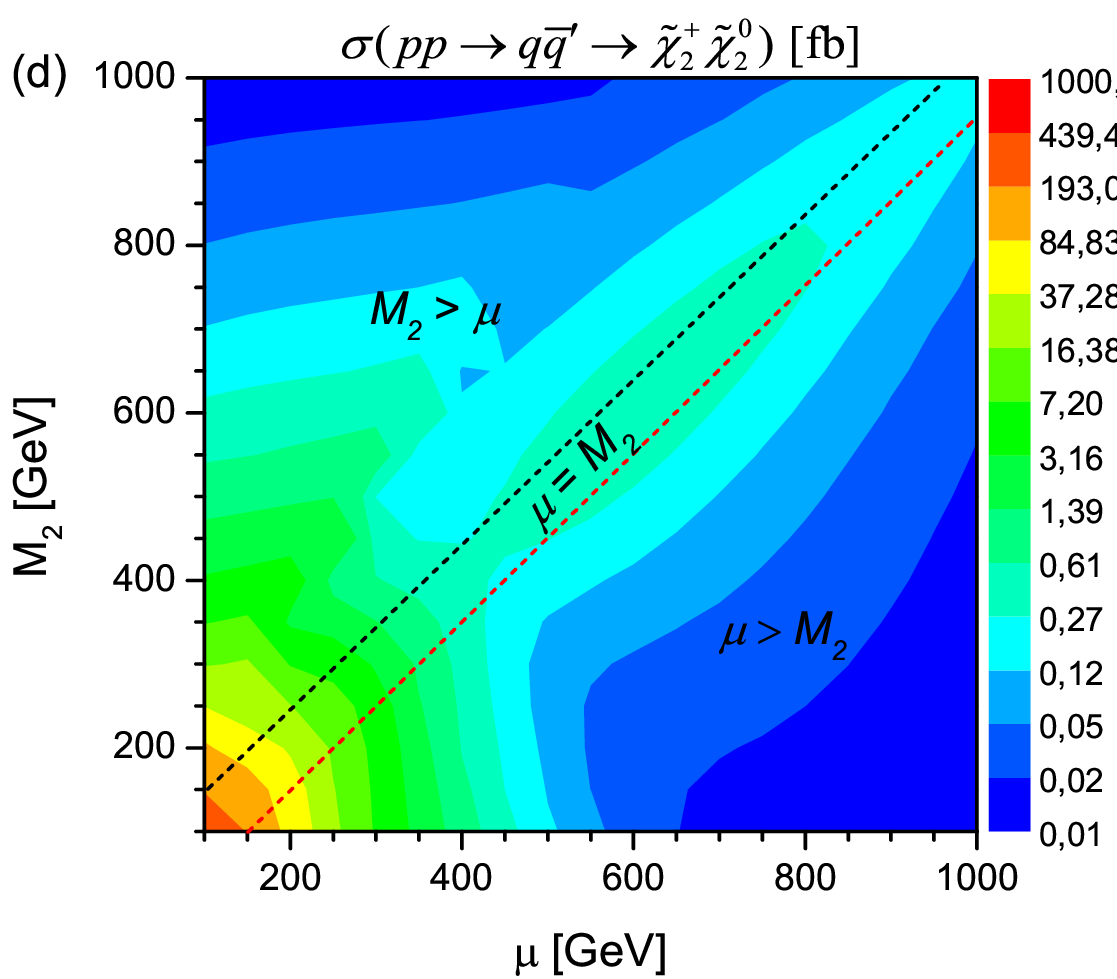}
      \end{center}
\caption{(color online). Contour plots of the total cross sections
of the process $pp\to\widetilde{\chi}_{i}^0\widetilde{\chi}_{j}^+$
(i,j=1,2) in the $M_2-\mu$ plane for $\sqrt{s}=8$ TeV. We choose
$\tan\beta=45$ and fix $M_1=\frac{5}{3}M_2 \tan^2\theta_W$.}
\label{Fig11}
\end{figure}

\begin{figure}[hpt]
    \begin{center}
\includegraphics[scale=0.35]{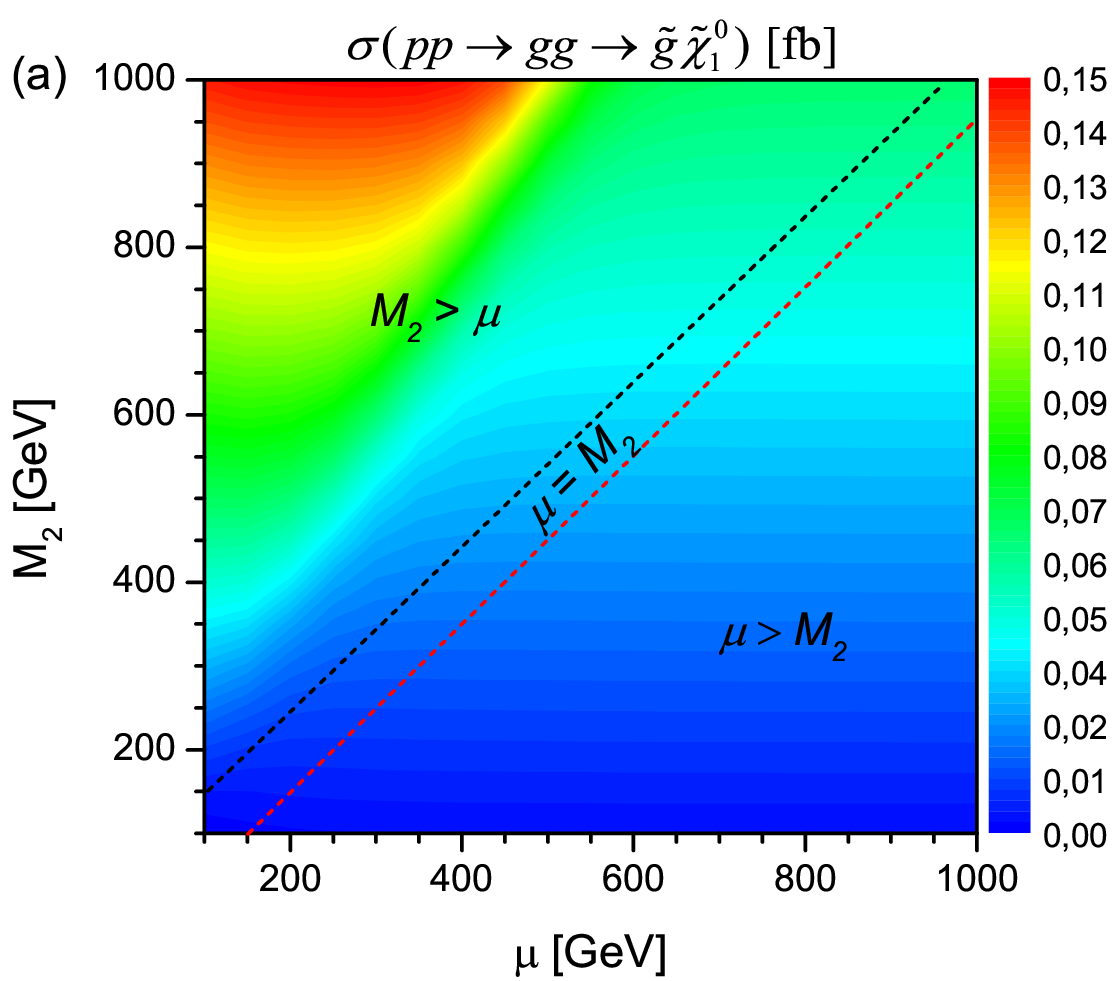}
\includegraphics[scale=0.35]{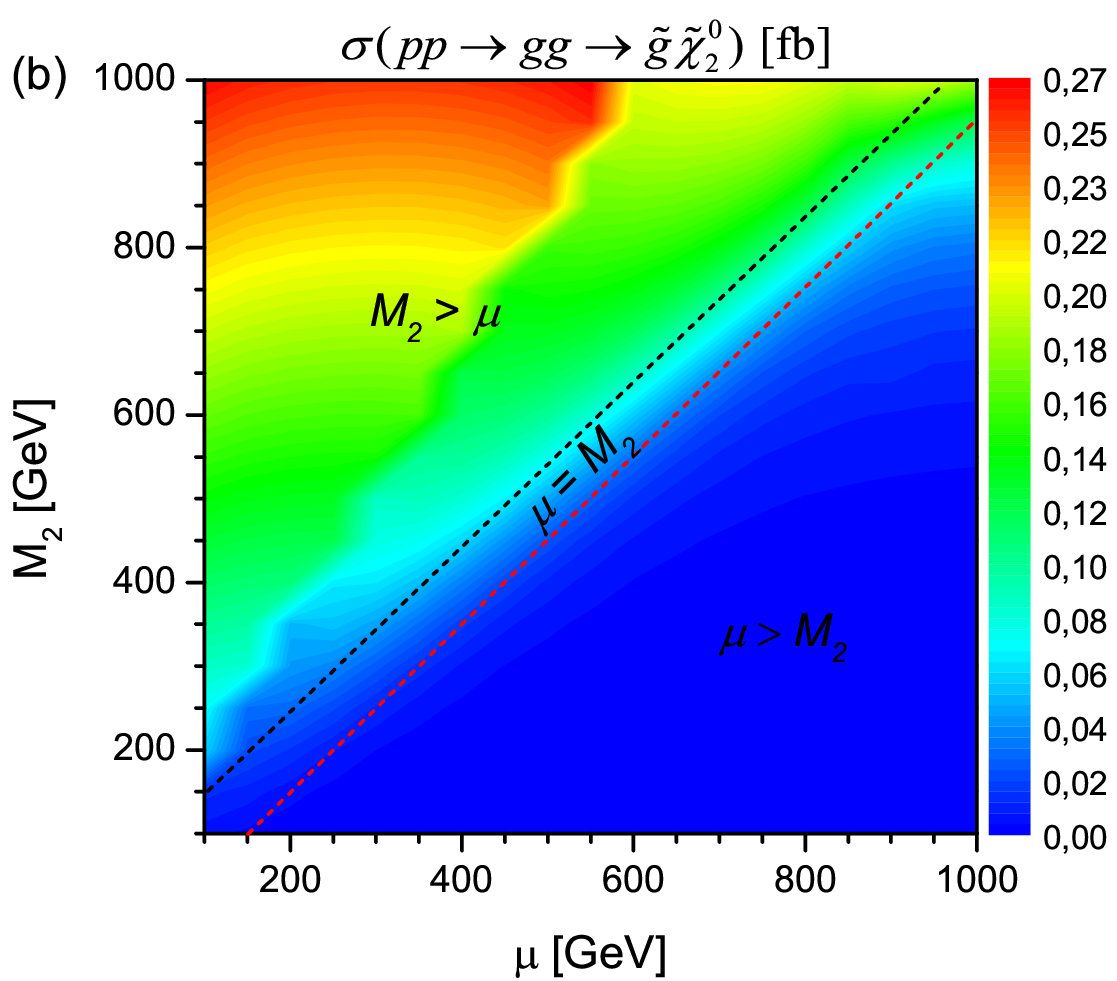}
     \end{center}
\caption{(color online). Contour plots of the total cross sections
of the process
$pp(\text{g}\text{g})\to\widetilde{\text{g}}\widetilde{\chi}_{i}^0$
(i=1,2) in the $M_2-\mu$ plane for $\sqrt{s}=8$ TeV. We choose
$\tan\beta=45$ and fix $M_1=\frac{5}{3}M_2 \tan^2\theta_W$.}
\label{Fig12}
\end{figure}

\begin{figure}[hpt]
    \begin{center}
\includegraphics[scale=0.36]{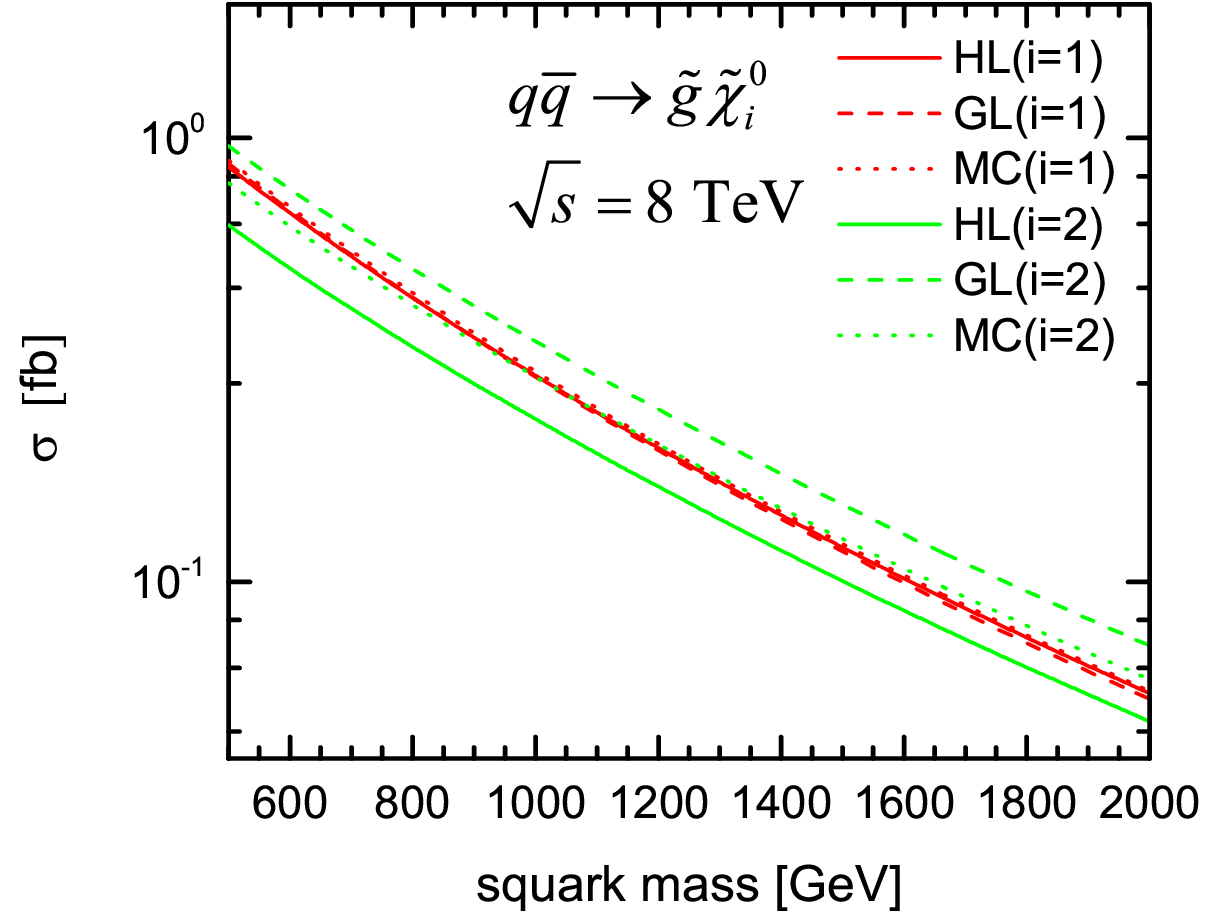}
     \end{center}
\caption{(color online). Total cross sections for the process
$pp\to\widetilde{\text{g}}\widetilde{\chi}_{i}^0$ (i=1,2) depending
on the squark mass at $\sqrt{s}=8$ TeV.} \label{Fig13}
\end{figure}

\begin{figure}[hpt]
    \begin{center}
\includegraphics[scale=0.36]{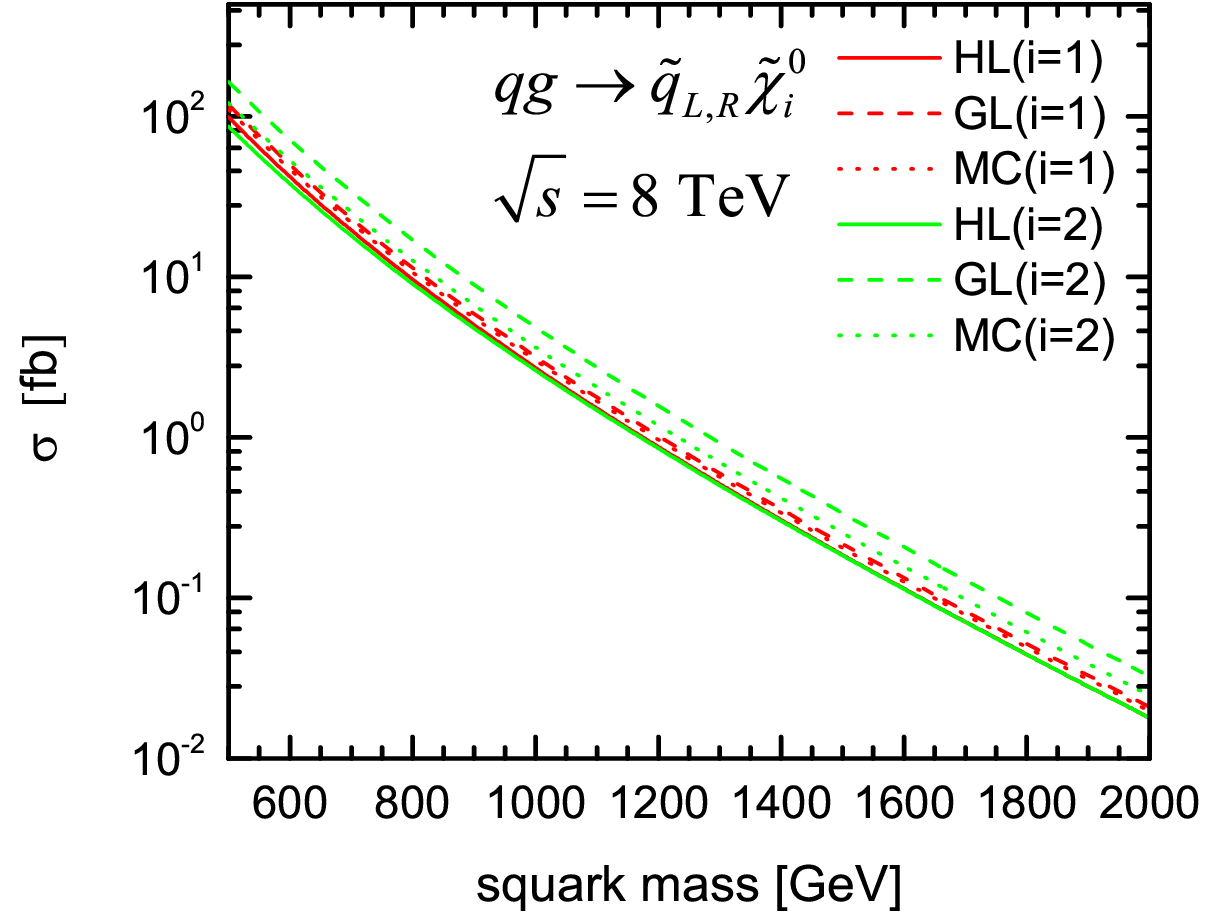}
     \end{center}
\caption{(color online). Total cross sections of the process
$pp\to\widetilde{q}_{L,R}\widetilde{\chi}_{i}^0$ (i=1,2) depending
on the squark mass at $\sqrt{s}=8$ TeV.} \label{Fig14}
\end{figure}

\begin{figure}[hpt]
    \begin{center}
\includegraphics[scale=0.36]{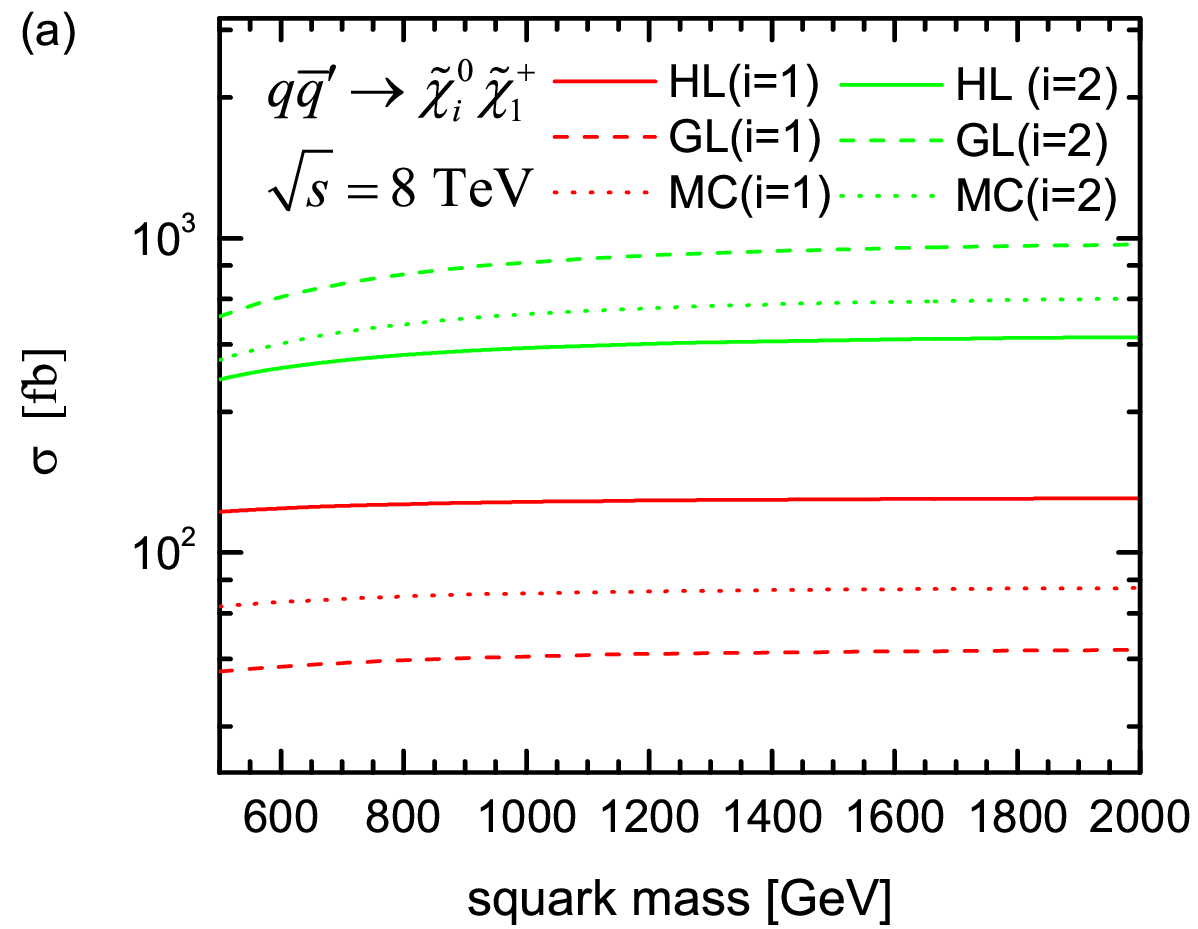}
\includegraphics[scale=0.36]{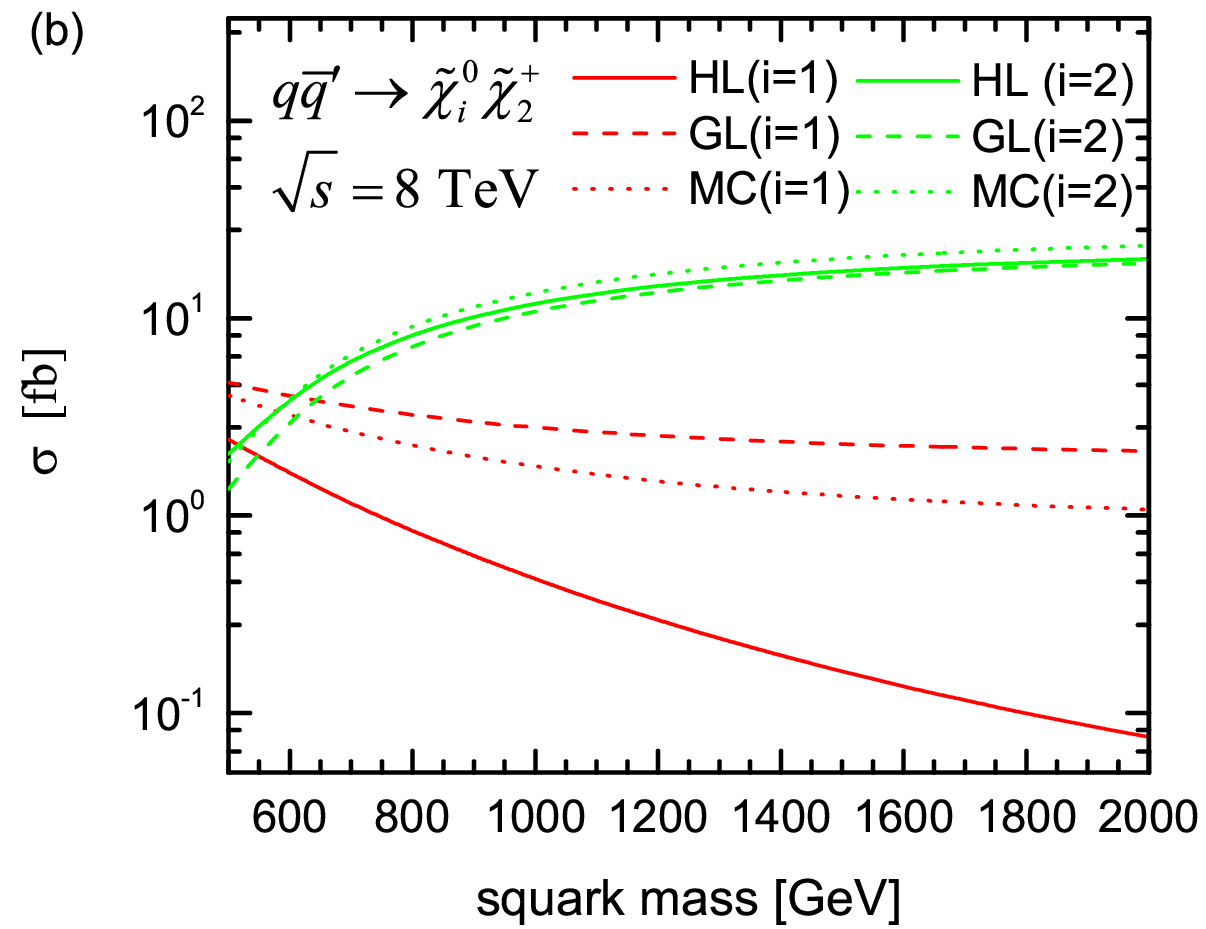}
     \end{center}
\caption{(color online). Total cross sections of the processes $pp
\to\widetilde{\chi}_{i}^0\widetilde{\chi}_{1}^+$ (left) and
$\widetilde{\chi}_{i}^0\widetilde{\chi}_{2}^+$ (right) (i=1,2)
depending on the squark mass at $\sqrt{s}=8$ TeV.} \label{Fig15}
\end{figure}

\begin{figure}[hpt]
    \begin{center}
\includegraphics[scale=0.36]{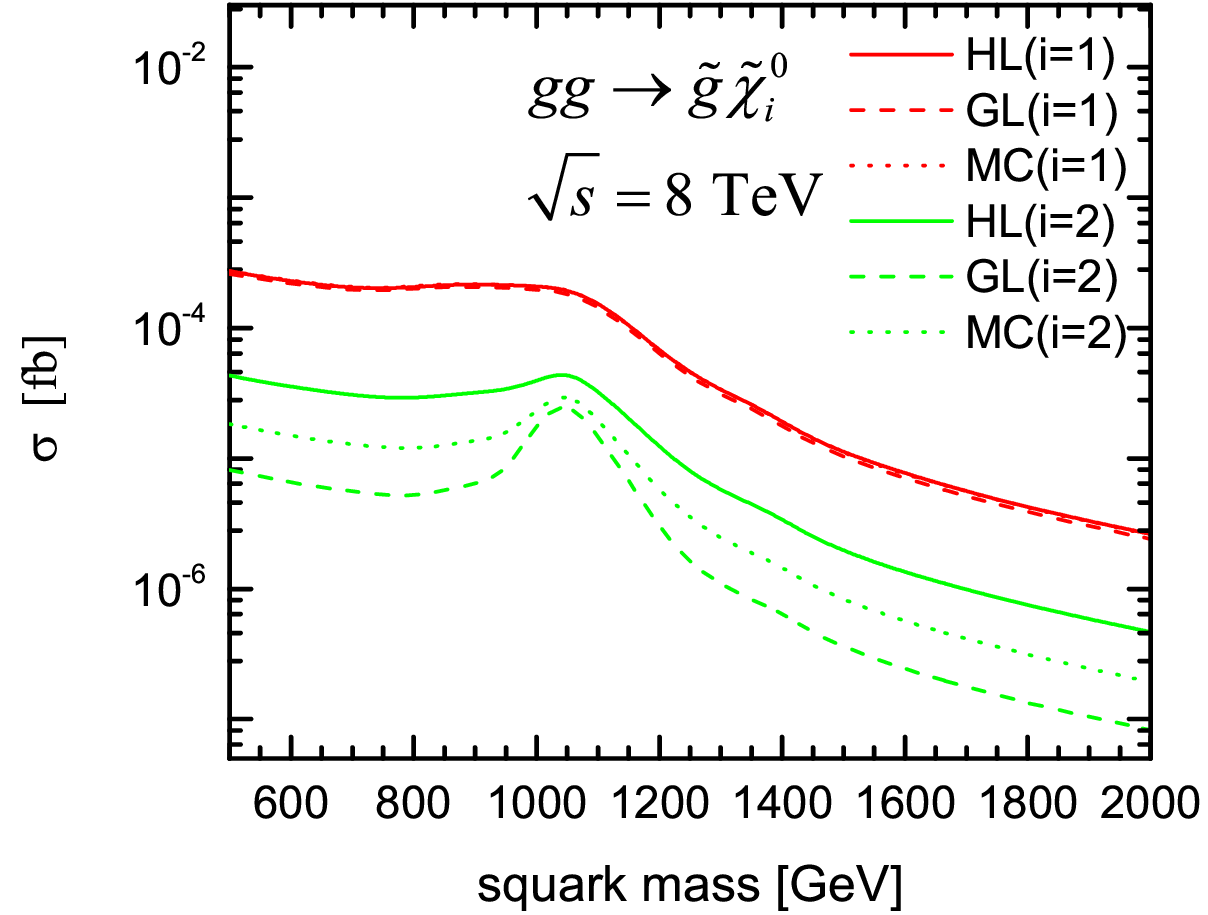}
     \end{center}
\caption{(color online). Total cross sections of the process
$pp(\text{g}\text{g})\to\widetilde{\text{g}}\widetilde{\chi}_{i}^0$
(i=1,2) depending on the squark mass at $\sqrt{s}=8$ TeV.}
\label{Fig16}
\end{figure}

\begin{figure}[hpt]
    \begin{center}
\includegraphics[scale=0.36]{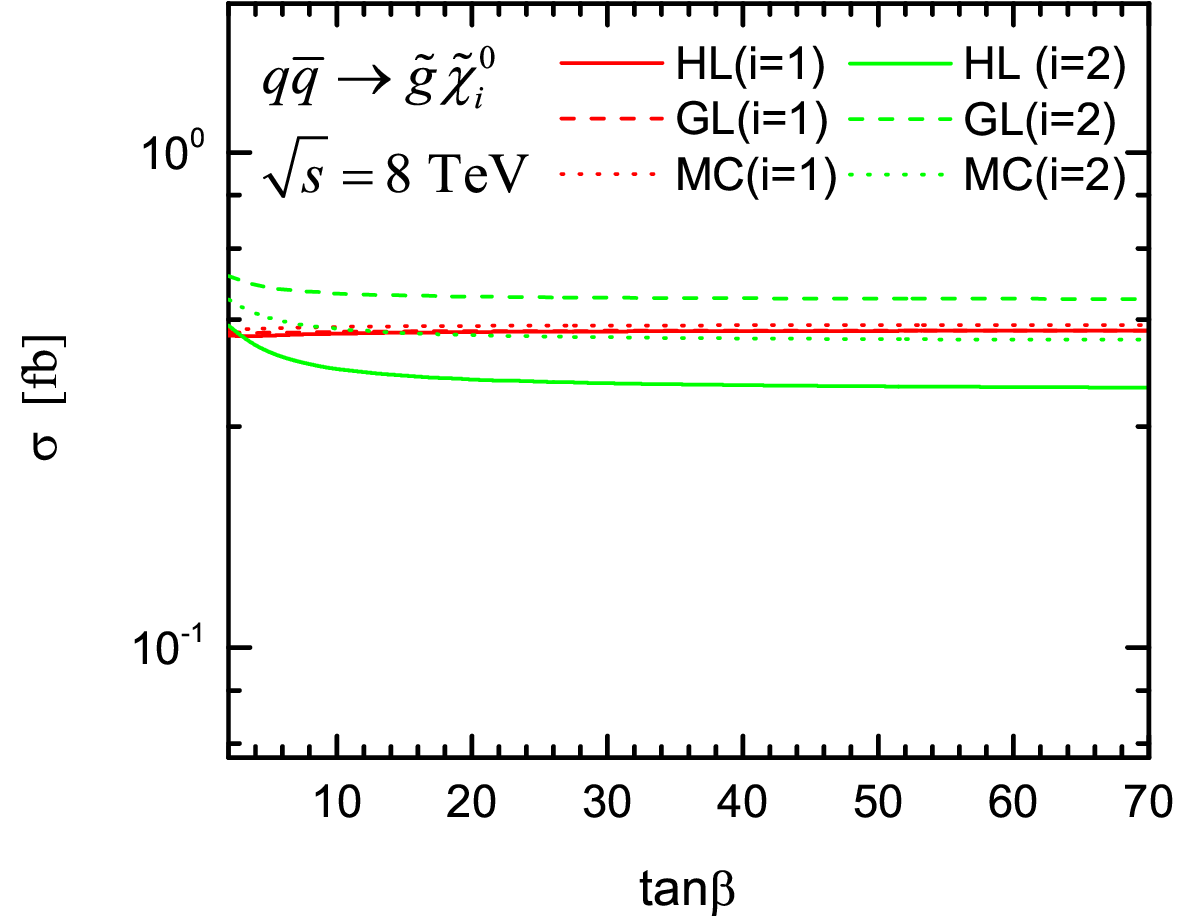}
     \end{center}
\caption{(color online). Total cross sections of the process
$pp\to\widetilde{\text{g}}\widetilde{\chi}_{i}^0$ (i=1,2) as a
function of $\tan{\beta}$ at $\sqrt{s}=8$ TeV.} \label{Fig17}
\end{figure}

\begin{figure}[hpt]
    \begin{center}
\includegraphics[scale=0.36]{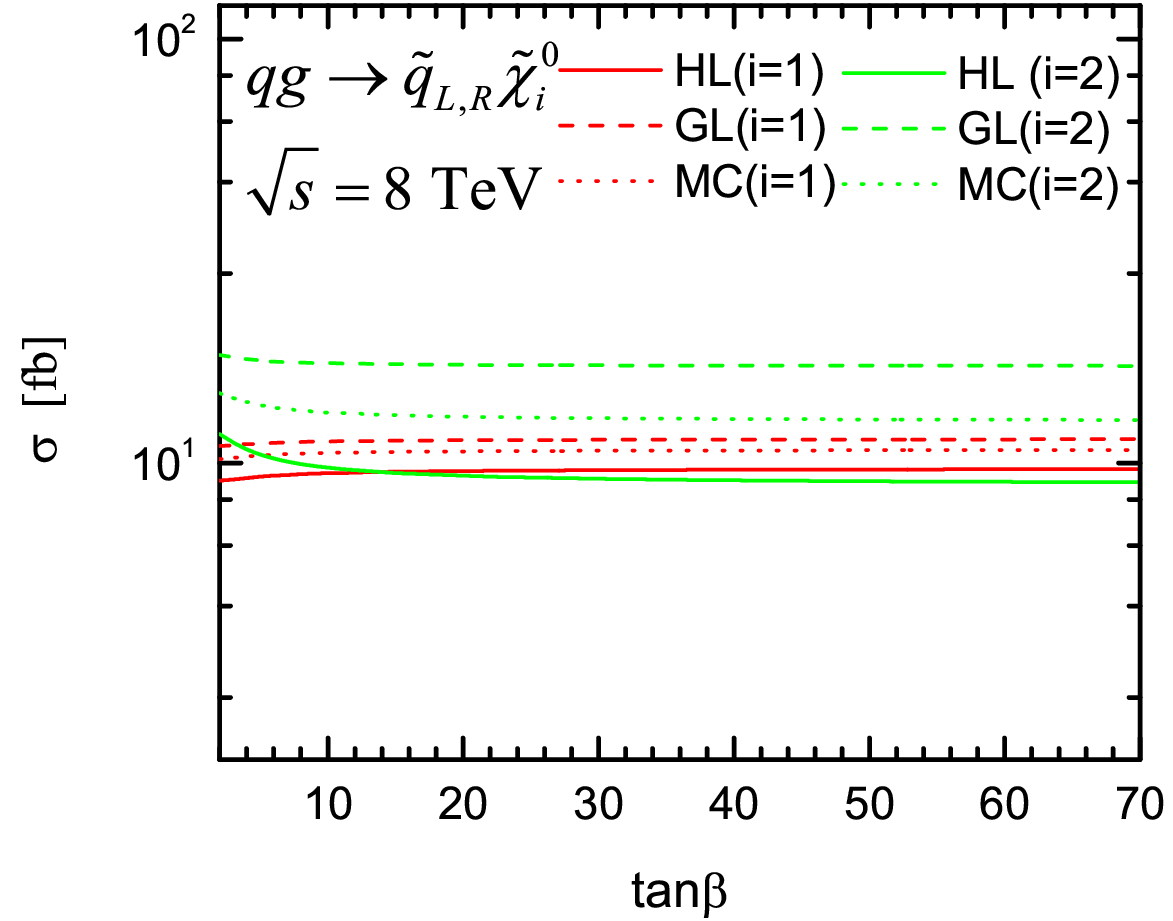}
     \end{center}
\caption{(color online). Total cross sections of the process
$pp\to\widetilde{q}_{L,R}\widetilde{\chi}_{i}^0$ (i=1,2) as a
function of $\tan{\beta}$ at $\sqrt{s}=8$ TeV.} \label{Fig18}
\end{figure}

\begin{figure}[hpt]
    \begin{center}
\includegraphics[scale=0.36]{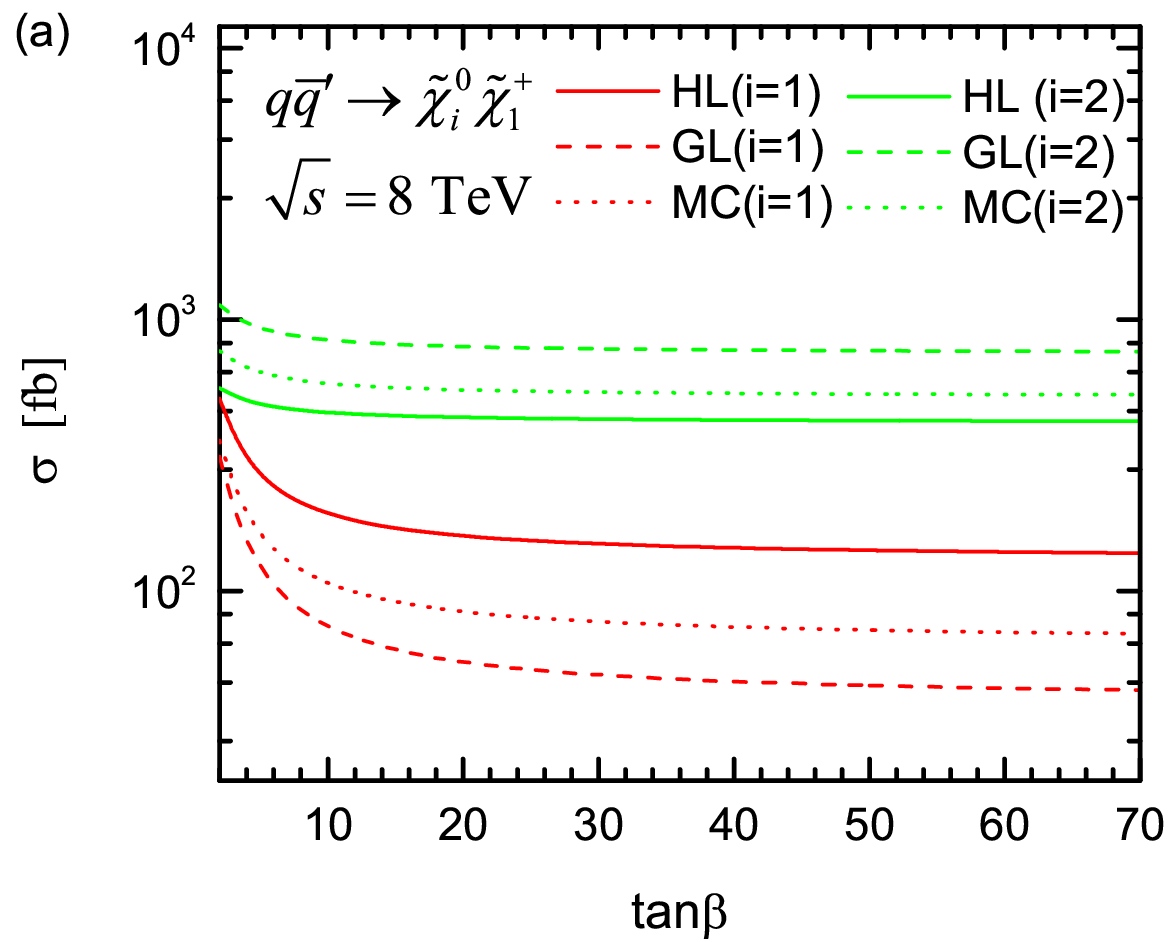}
\includegraphics[scale=0.36]{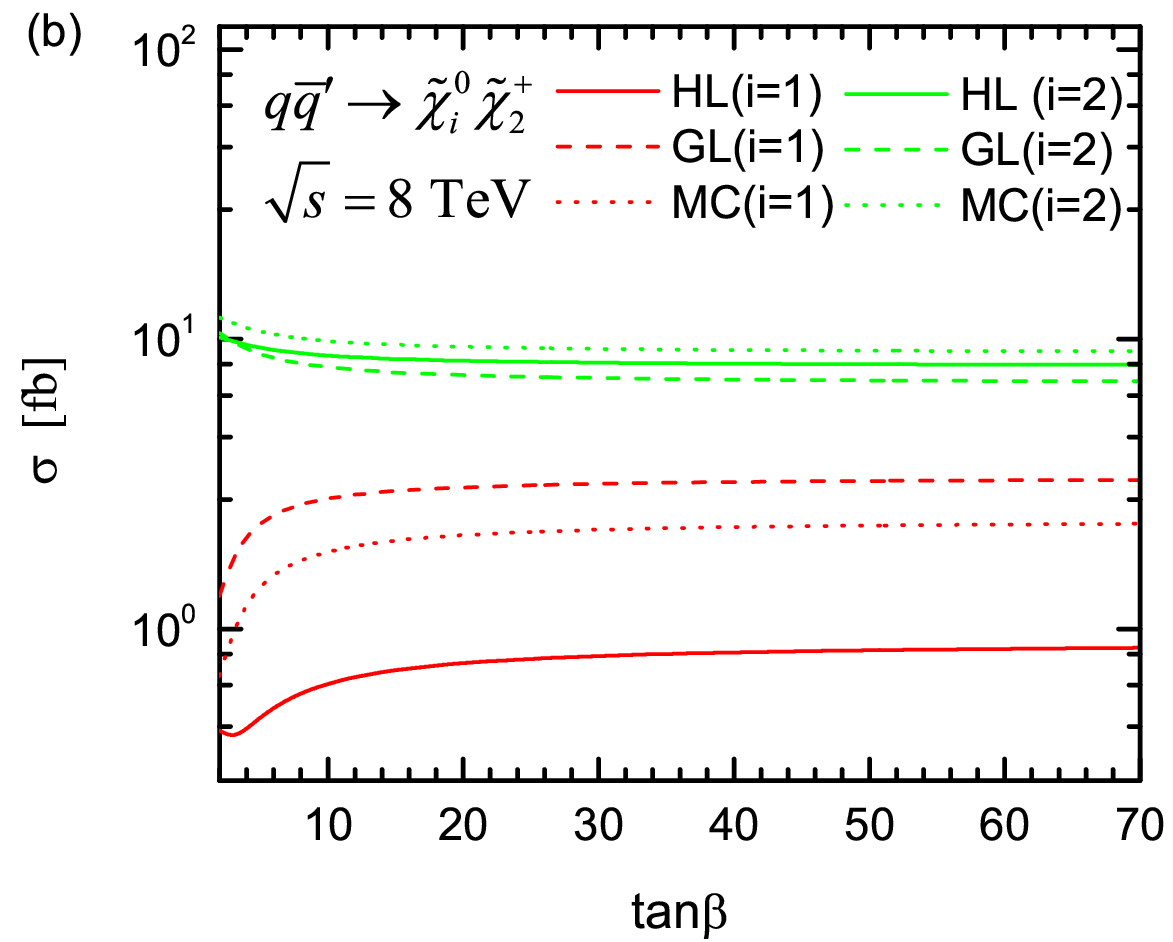}
      \end{center}
\caption{(color online). Total cross sections of the processes $pp
\to\widetilde{\chi}_{i}^0\widetilde{\chi}_{1}^+$ (left) and
$\widetilde{\chi}_{i}^0\widetilde{\chi}_{2}^+$ (right) (i=1,2) as a
function of $\tan{\beta}$ at $\sqrt{s}=8$ TeV.} \label{Fig19}
\end{figure}

\begin{figure}[hpt]
    \begin{center}
\includegraphics[scale=0.36]{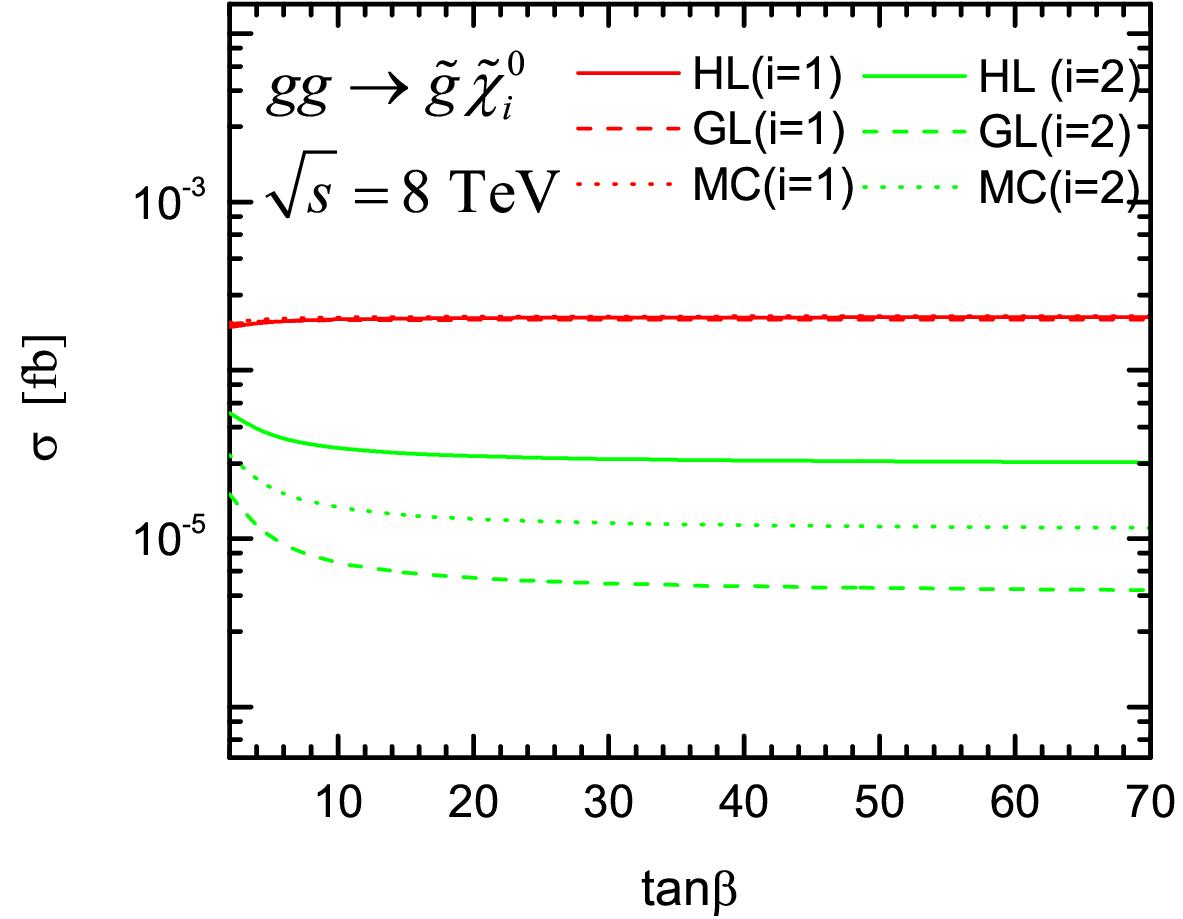}
     \end{center}
\caption{(color online). Total cross sections of the process
$pp(\text{g}\text{g})\to\widetilde{\text{g}}\widetilde{\chi}_{i}^0$
(i=1,2) as a function of $\tan{\beta}$ at $\sqrt{s}=8$ TeV.} \label{Fig20}
\end{figure}

We use the MSTW2008 parton distribution functions \cite{MSTW} for
the quark/gluon distributions inside the proton and fix the
renormalization and factorization scales to the average final-state
mass in our numerical calculations. For each scenario given above,
we have numerically evaluated the hadronic cross sections of the
single neutralino production processes involving a neutralino
$\widetilde{\chi}_{1}^{0}$ or $\widetilde{\chi}_{2}^{0}$ in the
final state, as a function of the center-of-mass energy from
Figs.~\ref{Fig5} to \ref{Fig8}, the $M_2$-$\mu$ mass plane from
Figs.~\ref{Fig9} to \ref{Fig12}, the squark mass from
Figs.~\ref{Fig13} to \ref{Fig16}, and $\tan\beta$ from
Figs.~\ref{Fig17} to \ref{Fig20}. In some of the figures, we use
abbreviations as follows: higgsino-like $\rightarrow$ HL(solid line), gaugino-like $\rightarrow$ GL(dashed line),
mixture-case $\rightarrow$ MC(dotted-line) and
CMSSM 40.2.2 benchmark point $\rightarrow$ CMSSM(dot-dashed line), respectively. We
now offer the following analysis of these figures in detail,
separately.

In~\cref{Fig5,Fig6,Fig7,Fig8}, we plot the dependence of the total
cross sections for the single neutralino processes of the
center-of-mass energy. These figures indicate that the total cross
sections increase slowly and smoothly with increasing the beam
energy from 7 TeV to 14 TeV for each scenario. The CMSSM 40.2.2
benchmark point and gaugino-like scenario are dominant for
$pp\rightarrow\widetilde\chi_{i}^{0}\widetilde{\text{g}}$ and $pp
\rightarrow\widetilde\chi_{i}^{0}\widetilde{q}_{L,R}$, respectively;
however, in the associated production of a chargino with
$\widetilde{\chi}_{i}^{0}$, these dominancies vary such that the
gaugino-like scenario is dominant for $pp \to
\widetilde{\chi}_{1}^{+}\widetilde{\chi}_{2}^{0}$ and
$\widetilde{\chi}_{2}^{+}\widetilde{\chi}_{1}^{0}$ while the
higgsino-like and mixture-case scenarios are dominant for $pp
\to\widetilde{\chi}_{1}^{+}\widetilde{\chi}_{1}^{0}$ and
$\widetilde{\chi}_{2}^{+}\widetilde{\chi}_{2}^{0}$ because of
contributions to cross section from not only neutralino mixing
matrix but also chargino mixing matrixes. The difference of the
cross sections in scenarios comes only from the change of the
couplings given in Eqs.~\eqref{eq:O_ij}-\eqref{eq:C_Cqsq} where the
mixing matrices are changed. For cross sections of the process
$\text{g} \text{g}\to \widetilde{\chi}_{i}^{0}\widetilde{\text{g}}$
at one-loop, higgsino-like scenario is larger than other scenarios.
As shown in Fig.~\ref{Fig5}, the cross section of the process $p p
\to \widetilde{\chi}_{1}^{0} \widetilde {\text{g}}$ in the CMSSM
40.2.2 benchmark point is about 9 times larger than in the
gaugino-like, higgsino-like and mixture-case scenarios. Also, the cross section of the process $p p \to
\widetilde{\chi}_{2}^{0} \widetilde {\text{g}}$ in the the CMSSM
40.2.2 benchmark point is 7, 9 and 11
times larger than in the gaugino-like, mixture-case and
higgsino-like scenarios, respectively. As seen from Fig.~\ref{Fig6}, the cross
section of the process $p p \to \widetilde{\chi}_{1}^{0} \widetilde
{q}_{L,R}$ in the gaugino-like scenarios is about 17\%, 6\%, and 4 times larger than in the
higgsino-like scenario, mixture-case scenario and CMSSM 40.2.2
benchmark point, respectively.
Also, the cross section of the process
$p p \to \widetilde{\chi}_{2}^{0} \widetilde {q}_{L,R}$ in the
gaugino-like scenario is 87\%,
34\% and 5 times larger than in the higgsino-like scenario,
mixture-case scenarios and CMSSM 40.2.2 benchmark point, respectively. It can be seen from in
Fig.~\ref{Fig7}(a) that the cross section of the process $p p \to
\widetilde{\chi}_{1}^{0}\widetilde{\chi}_{1}^{+} $ in the
higgsino-like scenario is 3.2 times,
96\% and 3 orders of magnitude larger than in the gaugino-like scenario,
mixture-case scenario and CMSSM 40.2.2 benchmark point, respectively. The cross
section of the process $p p \to
\widetilde{\chi}_{2}^{0}\widetilde{\chi}_{1}^{+} $ in the
gaugino-like scenario is roughly 2
times, 44\% and 1 orders of magnitude larger than in the higgsino-like scenario,
mixture-case scenario and CMSSM 40.2.2 benchmark point, respectively. Also, as
shown in Fig.~\ref{Fig7}(b), the cross section of the process $p p
\to \widetilde{\chi}_{1}^{0} \widetilde{\chi}_{2}^{+} $ in the
gaugino-like scenario is roughly 3.6 times, 1.4 times and 1 orders of magnitude larger than in the higgsino-like scenario, mixture-case scenario and CMSSM 40.2.2 benchmark point, respectively. The cross
section of the process $p p \to \widetilde{\chi}_{2}^{0}
\widetilde{\chi}_{2}^{+} $ in the
 mixture-case scenario is roughly 11\%, 27\% and 3 orders of magnitude larger than in the higgsino-like scenario, in the gaugino-like scenario and the CMSSM 40.2.2 benchmark point, respectively. As
shown in Fig.~\ref{Fig8}, the cross section of the process $\text{g}
\text{g}\to \widetilde{\chi}_{1}^{0}\widetilde{\text{g}}$ in the CMSSM 40.2.2 benchmark point is about 7.4, 7.1 and 7 times larger than in the gaugino-like scenario, higgsino-like scenario and mixture-case scenario, respectively. The cross section for
$\text{g} \text{g}\to \widetilde{\chi}_{2}^{0}\widetilde{\text{g}}$ in the CMSSM 40.2.2 benchmark point
is about 6.7 times, 2.8 times and 10\% larger than in the gaugino-like scenario, mixture-case scenario and higgsino-like scenario, respectively.
\begin{table}[htp]
\caption{Total cross sections (in fb) for the single neutralino
production at center-of-mass energy $\sqrt s=$ 7 and 14
TeV.}\label{table2}
\begin{ruledtabular}
\begin{tabular}{lrrrrrrrr}
\multirow{1}{*}{} &\multicolumn{2}{c}{Higgsino like}&\multicolumn{2}{c}{Gaugino like}&\multicolumn{2}{c}{Mixture case}&\multicolumn{2}{c}{CMSSM 40.2.2}\\ \cline{2-9}
$\sigma$(process) [fb]&~7 TeV &14 TeV& ~7 TeV &14 TeV& ~7 TeV &14 TeV&~7 TeV &14
TeV\\\hline
$\sigma(pp\to\widetilde{\chi}_{1}^{0}\widetilde{\text{g}})$&0.22&3.70 &0.22 &3.61 &0.23&3.75 &3.66&22.44\\
$\sigma(pp\to\widetilde{\chi}_{2}^{0}\widetilde{\text{g}})$&0.17&3.13&0.25 &4.80&0.21&3.97&3.15&24.79\\
 \noalign{\smallskip}\noalign{\smallskip}
$\sigma(pp\to\widetilde{\chi}_{1}^{0}\widetilde{q}_{L,R})$ &6.07 &48.66 &7.18 &56.59&6.75 &53.67&1.11&16.07\\
$\sigma(pp\to\widetilde{\chi}_{2}^{0}\widetilde{q}_{L,R})$ &5.63&47.82&10.50 &89.95&7.84&67.33&1.22&23.31\\
\noalign{\smallskip}\noalign{\smallskip}
$\sigma(pp\to\widetilde{\chi}_{1}^{0}\widetilde{\chi}_{1}^{+})~~$&117.92&296.75&37.83&93.53&60.10&150.75&0.24&0.80\\
$\sigma(pp\to\widetilde{\chi}_{2}^{0}\widetilde{\chi}_{1}^{+})~~$&346.94&922.03&629.53&1654.78&434.16&1157.26&59.10&163.32\\
$\sigma(pp\to\widetilde{\chi}_{1}^{0}\widetilde{\chi}_{2}^{+})~~$&0.64&2.24&2.56 &7.67&1.78&5.52&0.06&0.22\\
$\sigma(pp\to\widetilde{\chi}_{2}^{0}\widetilde{\chi}_{2}^{+})~~$&6.54 &19.11&5.78&16.64&7.31&21.11&0.01&0.04\\
\noalign{\smallskip}\noalign{\smallskip}
$\sigma(pp\to\widetilde{\chi}_{1}^{0}\widetilde{\text{g}})_{\text{one-loop}}$&$\mathscr{O}(10^{-4})$&$\mathscr{O}(10^{-3})$&$\mathscr{O}(10^{-4})$&$\mathscr{O}(10^{-3})$&$\mathscr{O}(10^{-4})$&$\mathscr{O}(10^{-3})$&$\mathscr{O}(10^{-3})$&0.02\\
$\sigma(pp\to\widetilde{\chi}_{2}^{0}\widetilde{\text{g}})_{\text{one-loop}}$&$\mathscr{O}(10^{-5})$&$\mathscr{O}(10^{-4})$&$\mathscr{O}(10^{-6})$&$\mathscr{O}(10^{-4})$&$\mathscr{O}(10^{-6})$&$\mathscr{O}(10^{-4})$&$\mathscr{O}(10^{-5})$&$\mathscr{O}(10^{-4})$\\
\end{tabular}
\end{ruledtabular}
\end{table}

In~\cref{table2}, the cross sections of single neutralino associated
production at center-of-mass energy $\sqrt s=$ 7 TeV and 14 TeV are
given for each scenario. It is clear from this table that the cross
section of the process
$pp\to\widetilde{\chi}_{2}^{0}\widetilde{\chi}_{1}^{+}$ in the
gaugino-like scenario yields cross sections of $\thicksim$600 to 1700
fb for $\sqrt{s}=$ 7 TeV and 14 TeV, which is larger than the
remaining ones. Moreover, the cross section for
$pp\to\widetilde{\chi}_{1}^{0}\widetilde{q}_{L,R}(\widetilde{\chi}_{2}^{0}\widetilde{q}_{L,R})$
reaches about 57(90) fb at  $\sqrt{s}=$ 14 TeV in the
gaugino-like. However, the process $pp(\text{g} \text{g})\to
\widetilde{\chi}_{i}^{0}\widetilde{\text{g}}$ is suppressed by the
others. The magnitudes of the cross sections are at a visible level of
$10^0$ fb for
$pp\rightarrow\widetilde\chi_{i}^{0}\widetilde{\text{g}}$, $10^1$ fb
for $pp \rightarrow\widetilde\chi_{i}^{0}\widetilde{q}_{L,R}$,
$10^{-1}$-$10^3$ fb for
$pp\rightarrow\widetilde\chi_{i}^{0}\widetilde\chi_{j}^{+}$, and
$10^{-4}$-$10^{-2}$ fb for
$pp(\text{g}\text{g})\rightarrow\widetilde\chi_{i}^{0}\widetilde{\text{g}}$
at $\sqrt{s}=$ 14 TeV. Additionally, it can be easily seen that the
cross section for the associated production of the next-to-lightest
neutralino $\widetilde{\chi}_{2}^{0}$ is generally much larger than
the cross section for associated production the lightest neutralino
$\widetilde{\chi}_{1}^{0}$ for each scenario.

The masses and mixing matrices of neutralino/chargino depend on the
parameters $M_2$ and $\mu$; therefore, it is so important to study
the dependence of the cross section of the single neutralino
production on these parameters. Accordingly, we plot the dependence
of the total cross section of the associated process in the
$M_2$-$\mu$ mass plane with varying $M_2$ and $\mu$ in the range
from 100 to 1000 GeV in steps of 50 GeV at center-of-mass energy 8
TeV for $\tan\beta=$ 45, as shown in~\cref{Fig9,Fig10,Fig11,Fig12}.
In these figures, the region above the black dashed-line corresponds to
$M_2>\mu$ (higgsino-like) , the region below the red dashed-line
corresponds to $M_2<\mu$ (gaugino-like) and the region between the two
dashed lines corresponds to $\mu=M_2$ (mixture-case). One can note
that these figures reconfirm the dominant scenarios which appear in
the dependence of the cross sections on the center-of-mass energy.
We can see from~\cref{Fig9,Fig10} that the cross sections of the
processes $pp\rightarrow\widetilde\chi_{i}^{0}\widetilde{\text{g}}$
and $pp \rightarrow\widetilde\chi_{i}^{0}\widetilde{q}_{L,R}$ in the
$M_2$-$\mu$ mass plane increase during both increasing $\mu$ and
decreasing $M_2$. In particular, the maximum values are obtained in
the region 200 $\lesssim\mu \lesssim$ 1000 GeV and $M_2 \lesssim$
400 GeV into the scan region. This case corresponds to the gaugino-like
scenario. As a result, one can note that the cross section of these
processes can be measured experimentally in some scenarios for a lower
value of $M_2$. However, as illustrated in Fig.~\ref{Fig11}, the
cross sections for $pp \to \widetilde
{\chi}_{i}^{0}\widetilde{\chi}_{j}^{+}$ in the $M_2$-$\mu$ mass
plane increase during both decreasing $\mu$ and $M_2$. Here, the
maximum values are obtained in the region $\mu \lesssim$ 400 GeV and
any value of $M_2$ for processes $pp \to \widetilde
{\chi}_{1}^{0}\widetilde{\chi}_{1}^{+}$ ($M_2>\mu$) and $\widetilde
{\chi}_{2}^{0}\widetilde{\chi}_{1}^{+}$ ($\mu>M_2$), while in the
region 100 $\lesssim M_2 \lesssim$ 400 GeV and 100 $\lesssim\mu
\lesssim$ 400 GeV for processes $pp \to
\widetilde{\chi}_{1}^{0}\widetilde{\chi}_{2}^{+}$ ($\mu
>M_2$) and $\widetilde{\chi}_{2}^{0}\widetilde{\chi}_{2}^{+}$
($\mu=M_2$). Note that, as mentioned before, the process of
contributions to cross section from not only neutralino mixing
matrix, but also chargino mixing matrixes. One can see from
Fig.~\ref{Fig12} that the dependence of the cross section of the
process $\text{g} \text{g}\to
\widetilde{\chi}_{i}^{0}\widetilde{\text{g}}$ in the $M_2$-$\mu$
mass plane increases with increasing $M_2$ and any value of $\mu$. In
particular, the cross section of process $\text{g} \text{g}\to
\widetilde{\chi}_{i}^{0}\widetilde{\text{g}}$ indicates the maximum
values in the region 600 $\lesssim M_2 \lesssim$ 1000 GeV and $\mu
\lesssim$ 600 GeV as illustrated in Figs.~\ref{Fig12}(a) and~\ref{Fig12}(b). This
case corresponds to higgsino-like scenario ($M_2>\mu$).

In~\cref{Fig13,Fig14,Fig15,Fig16} we present the cross section as a
function of squark mass for single neutralino production at
$\sqrt{s}=$ 8 TeV. The total cross section for the single neutralino
production processes apart from $pp \to \widetilde
{\chi}_{i}^{0}\widetilde{\chi}_{j}^{+}$ are essentially determined
by the squark masses so that it decreases with increasing the
squark mass between 500 and 2000 GeV for each scenario. When the
squark mass increases by a factor of 4, the cross section is pulled
down by about 1, 3 and 2 orders of magnitude for the processes
$pp\rightarrow\widetilde\chi_{i}^{0}\widetilde{\text{g}}$, $pp
\rightarrow\widetilde\chi_{i}^{0}\widetilde{q}_{L,R}$ and
$pp(\text{g} \text{g})\to
\widetilde{\chi}_{i}^{0}\widetilde{\text{g}}$, respectively. On the
other hand, for the process $pp \to \widetilde
{\chi}_{i}^{0}\widetilde{\chi}_{j}^{+}$, the cross section is less
affected with respect to variation in the squark mass because the
\textit{s}-channel of this process is dominant and together \textit{t}- and \textit{u}-channel
terms are suppressed for large squark masses. The cross sections of
the single neutralino production for the squark mass 1 and 2 TeV at
$\sqrt{s}=$ 8 TeV so as to facilitate precise comparisons with the
experimental results are summarized in~\cref{table3}.
As seen from this table, the dependence of cross section on the
squark mass is dominated by one of the processes, $pp \to \widetilde
{\chi}_{2}^{0}\widetilde{\chi}_{1}^{+}$ appears 0.95 pb for the
squark mass 2 TeV in the gaugino-like scenario.

\begin{table*}[htp]
\caption{Total cross sections (in fb) for the single neutralino
production processes in a function of the squark mass at $\sqrt{s}=$
8 TeV.}\label{table3}
\begin{ruledtabular}
\begin{tabular}{lcccrrrcccrc}
~~~~&$m_{\widetilde{q}}$~[GeV]~~~&~$\widetilde{\chi}_{1}^{0}\widetilde{\text{g}}$~~&$\widetilde{\chi}_{2}^{0}\widetilde{\text{g}}$&~~~$\widetilde{\chi}_{1}^{0}\widetilde{q}$&$\widetilde{\chi}_{2}^{0}\widetilde{q}$&~~$\widetilde{\chi}_{1}^{0}\widetilde{\chi}_{1}^{+}$~~&$\widetilde{\chi}_{2}^{0}\widetilde{\chi}_{1}^{+}$
&$\widetilde{\chi}_{1}^{0}\widetilde{\chi}_{2}^{+}$&$\widetilde{\chi}_{2}^{0}\widetilde{\chi}_{2}^{+}$&$~\widetilde{\chi}_{1}^{0}\widetilde{\text{g}}_{\text{one-loop}}$&$\widetilde{\chi}_{2}^{0}\widetilde{\text{g}}_{\text{one-loop}}$\\
\hline
\multirow{2}*{HL} &1000&0.29&0.23&2.72 &2.62 &144.90&447.40&0.48 &11.85&2.07$\cdot10^{-5}$&4.16$\cdot10^{-6}$\\
   &2000&0.06&0.05&0.02&0.02&148.73&484.43 &0.08&19.92&2.61$\cdot10^{-7}$&4.93$\cdot10^{-8}$\\
\multirow{2}*{GL} &1000&0.29&0.35&3.19 &4.87&46.78&838.21&2.79&10.86&1.95$\cdot10^{-5}$ &1.96$\cdot10^{-6}$\\
   &2000&0.05&0.07&0.02&0.03&49.01&954.97&2.13 &18.99&2.36$\cdot10^{-7}$ &0.86$\cdot10^{-8}$\\
\multirow{2}*{MC} &1000&0.30&0.29&3.01 &3.65 &74.26&573.41&1.77 &13.53&2.08$\cdot10^{-5}$ &2.45$\cdot10^{-6}$\\
   &2000&0.06&0.06&0.02&0.03&77.10&642.44&1.07 &23.31&2.57$\cdot10^{-7}$ &2.02$\cdot10^{-8}$\\
\end{tabular}
\end{ruledtabular}
\end{table*}

Finally, the $\tan\beta$ dependence of the cross sections for the single neutralino
processes are depicted in~\cref{Fig17,Fig18,Fig19,Fig20}. From these figures we can clearly see that cross sections of the processes $pp \to \widetilde {\chi}_{i}^{0}\widetilde{\text{g}}$,
$pp\to \widetilde {\chi}_{i}^{0}\widetilde{q}_{L,R}$ and $\text{g}
\text{g}\to \widetilde{\chi}_{i}^{0}\widetilde{\text{g}}$ increase (decrease) slowly for $i=1$ ($i=2$) when $\tan \beta$ goes up from 2 to 10, and vary smoothly  when $\tan \beta >$ 10 for each scenario.
However, the cross sections of the processes $pp \to
\widetilde {\chi}_{i}^{0}\widetilde{\chi}_{j}^{+}$ apart from $pp \to
\widetilde {\chi}_{1}^{0}\widetilde{\chi}_{2}^{+}$ decrease with increasing the $\tan \beta$ from 2 to 70.
Moreover, there appear the same dominant scenarios as in the dependence of the cross
sections on the center-of-mass energy.

The possible contributions to the background in the signal regions
come from the Standard Model processes, as $pp\to WW$,
$pp\to ZZ$, $pp\to WZ$ and $pp\to t \bar t$. If we are interested in signals with
leptons in the final state, then in the case of $ 1l+\slashed{E}_{T}+jets$ mode, the
background appears from $pp\to WW$, $pp\to WZ$ and $pp\to t\bar t$.
Also, the processes $pp\to ZZ$, $pp\to WW$, and $pp\to t\bar t$ can
yield background for the $2l+\slashed{E}_{T}+jets$ mode. The process $pp\to
WZ$ can yield background for the $3l+\slashed{E}_{T}+0jets$  decay mode. Of
course, all background channels could have large cross sections, but despite this
it needs some additional cutoff mechanism that will help for
the extraction, as mentioned above. An analysis of our calculations is
shown since those background channels can have large cross
sections. It should be noted that, in our case, the background
cross section is about 1-3 orders of magnitude larger than the
signal. We hope the at $\sqrt s=$ 14 TeV with integrated luminosity
$L_{int}=100$ fb$^{-1}$, total cross section of single neutralino
production in the gaugino-like case could be observable at the
LHC. It should be noted that some problems within $E_{6}$ model are
discussed in Ref.~\cite{Frank}

\newpage
\section{Conclusion}\label{Conc}

In the present paper, we have concentrated on the single neutralino
production processes
$pp\rightarrow\widetilde\chi_{i}^{0}\widetilde{\text{g}}$, $pp
\rightarrow\widetilde\chi_{i}^{0}\widetilde{q}_{L,R}$,
$pp\rightarrow\widetilde\chi_{i}^{0}\widetilde\chi_{j}^{+}$ at
tree level and one loop
$pp(\text{g}\text{g})\rightarrow\widetilde\chi_{i}^{0}\widetilde{\text{g}}$
at the LHC. Cross sections of these processes have been calculated by the
CMSSM 40.2.2 benchmark point and three different scenarios as named
higgsino-like, gaugino-like and mixture cases. From our calculations,
we have obtained that in this cases, the gaugino-like scenario was more dominant
relative to the other scenarios. Additionally, the processes
$pp\rightarrow\widetilde\chi_{1}^{0}\widetilde\chi_{1}^{+}$ and
$\widetilde\chi_{2}^{0}\widetilde\chi_{1}^{+}$ dominated over the
other single neutralino production processes by roughly 2-3 orders
of magnitude. In particular, the cross section of the process $p p
\to \widetilde{\chi}_{2}^{0}\widetilde{\chi}_{1}^{+}$ in the
gaugino-like scenario ($
\widetilde{\chi}_{1}^{0}\widetilde{\chi}_{1}^{+}$ in the
higgsino-like scenario) appeared in the range of $\thicksim$0.63 (0.12)
pb to $\thicksim$1.65 (0.30) pb with increasing centre-of-mass energy
from 7 to 14 TeV. One may argue that the investigation of these
two processes for the single neutralino production at proton-proton
collisions is significant in both experimental and theoretical research.
According to our opinion, these may be used as a probe for
an experimental search on the single neutralino production in the
LHC and also in the future colliders. It is clear that the results
discussed in the parameter scan depend strongly on the assumptions
take into consideration, like the $M_2$ and $\mu$ parameters. The CMSSM
scenario have different character, which is more like the higgsino-like and mixture
cases. In general, our scenarios dominate over the CMSSM 40.2.2
benchmark scenario. Thus, taking into account the predictions of our
study in the LHC, single neutralino production processes are more
likely to be observed. Observables should then be constructed
addressing gluino, squark and neutralino decay channels to various
numbers of leptons and jets; as such, the $\widetilde{q} \to q
\widetilde{\chi}_{1}^{0}$ and $\widetilde{g} \to q \bar q
\widetilde{\chi}_{1}^{0}$ cascade decays to weakly interacting
neutralino which escape the detector unseen. Also, we hope our
results will help explain the expectation results in the LHC and
future linear collider.

\section*{Acknowledgments}
This work is supported by TUBITAK under grant number 2221(Turkey). One of the authors  A. I.~Ahmadov is grateful for financial support Baku State University Grant ``50+50''. Authors acknowledge
interest of members of the Department of Physics of Karadeniz
Technical University.

\appendix
\section{The neutralino/chargino sector of the MSSM}\label{Appendix}
The neutralino mass eigenstates $\widetilde\chi_i^{0}$ ($i=1,..,4$)
are the linear superposition of the gauginos $\widetilde B$,
$\widetilde W^{3}$  and the Higgsinos $\widetilde H_1^{0}$,
$\widetilde H_2^{0}$ in the MSSM. The neutralinos mass term in the
MSSM Lagrangian is expressed as \cite{Haber}
\begin{equation} \label{eq:Lagrangian}
\mathcal{L} = -\frac{1}{2}(\psi^0_i)^{T} \mathcal{M} \psi^0_j +h.c.,
\end{equation}
which is bilinear in the fermion fields ${\psi}_{j}^{0}=(-i
\widetilde B,-i \widetilde W^3,\widetilde H_1^{0},\widetilde
H_2^{0})^T$ with $j=1,..,4$. The neutralino mass matrix, which
is generally a complex and symmetric matrix, is explicitly given by
\begin{equation} \label{eq:MassMatrixN}
\mathcal{M}=\left(\begin{array}{cccc}M_{1}&0&-m_{Z}c_{\beta}s_{W}&m_{Z}s_{\beta}s_{W}\\
0&M_{2}&m_{Z}c_{\beta}c_{W}&-m_{Z}s_{\beta}c_{W}\\
-m_{Z}c_{\beta}s_{W}&m_{Z}c_{\beta}c_{W}&0&-\mu\\
m_{Z}s_{\beta}s_{W}&-m_{Z}s_{\beta}c_{W}&-\mu&0\end{array}\right),
\end{equation}
where $M_{1}$ and $M_{2}$ are the gaugino mass parameters
corresponding to the $U(1)$ and $SU(2)$ subgroups, separately,
$\mu$ is the Higgsino mass parameter, and $tan\beta =v_2/v_1$ equal
to the ratio of the vacuum expectation values $v_{1,2}$ of the two
Higgs doublets, which break the electroweak symmetry. These mass
parameters are complex in CP-noninvariant theories. The mass parameter $M_{2}$ could be achieved by the
reparametrization of the fields as real and
positive without any loss of generality so that the two remaining
nontrivial phases, which are reparametrization invariant, could be associated with
$M_{1}$ and $\mu$ as follows: $ M_{1}
=|M_{1}|e^{i\phi_{1}}$ and  $\mu=|\mu|e^{i\phi_{\mu}}, (0\leq\phi_1,\phi_\mu<2\pi)$.

The neutralino mass matrix $\mathcal{M}$  is diagonalized by a
$4\times4$ unitary matrix $N$, which is adequate to transform from the
gauge eigenstate basis ($\widetilde{B},\widetilde{W}^3,\widetilde
{H}_{1}^0,\widetilde {H}_{2}^0$) to the mass eigenstate basis of the
Majorana fields $\widetilde{\chi}_{i}^{0}$ such that,
\begin{equation} \label{eq:Md}
\mathcal{M}_D=N^{T}\mathcal{M}N=\sum_{j=1}^{4}m_{\widetilde{\chi}_{j}^{0}}E_{j}.
\end{equation}
The relation between the weak and physical neutralinos' eigenstates is expressed by $\chi_{i}^{0}=N_{ij}{\psi}_{j}^{0}$. For determining of the mixing matrix $N$, we get the square of the Eq.~\eqref{eq:Md} as follows:
\begin{equation} \label{eq:Md2}
\mathcal{M}_{D}^2=N^{-1}\mathcal{M}^{+}\mathcal{M}N=\sum_{j=1}^{4}m_{\widetilde{\chi}_{j}^{0}}^2E_{j},
\end{equation}
where $(E_j)_{ik}=\delta_{ji}\delta_{jk}$. The neutralino mass
eigenvalues $m_{\widetilde{\chi}_{j}^{0}}$ in $\mathcal{M}_{D}$ could
be gotten as reel and positive by an appropriate definition of the
unitary matrix $N$. From Eq.~\eqref{eq:Md2}, we get
\begin{equation} \label{eq:MM}
(\mathcal{M}^{+}\mathcal{M})N-N\mathcal{M}_{D}^2=0,
\end{equation}
and then considering the following relation
\begin{equation}
|N_{1j}|^2+|N_{2j}|^2+|N_{3j}|^2+|N_{4j}|^2=1,
\end{equation}
the unitary matrix $N_{ij}$ is determined from the system of equations in Eq.~\eqref{eq:MM} (see Ref.~\cite{Ahmadov} for details). Moreover, the neutralino masses are
solutions of the characteristic equation related to this system, which
is
\begin{equation}\label{eq:charac}
X^4-aX^3+bX^2-cX+d=0.
\end{equation}
After solution Eq.~\eqref{eq:charac}, one is able to get the exact
analytical expressions for the neutralino masses as follows:
\begin{equation} \label{eq:mN}
\begin{split}
  & m_{\widetilde{\chi}_{1}^{0}}^2,
m_{\widetilde{\chi}_{2}^{0}}^2=\frac{a}{4}-\frac{f}{2}\mp\frac{1}{2}\sqrt{r-w-\frac{p}{4f}}, \\
  & m_{\widetilde{\chi}_{3}^{0}}^2,
m_{\widetilde{\chi}_{4}^{0}}^2=\frac{a}{4}+\frac{f}{2}\mp\frac{1}{2}\sqrt{r-w+\frac{p}{4f}},
\end{split}
\end{equation}
where
\begin{equation}
\begin{split}
  & f=\sqrt{\frac
{r}{2}+w},~~r=\frac{a^2}{2}-\frac{4b}{3},~~w=\frac{q}{(3\cdot
2^{1/3})}+\frac{(2^{1/3}\cdot h)}{3\cdot q}, \\
  & p=a^3-4ab+8c,~~q=(k+{\sqrt{k^2-4h^3}})^{1/3}, \\
  & k=2b^3-9abc+27c^2+27a^2d-72bd,~~h=b^2-3ac+12d.
\end{split}
\end{equation}

The chargino mass eigenstates $\widetilde\chi_j^{\pm}$
($j=1,2$) are the linear superposition of the gauginos $\widetilde
W^{\pm}$ and the Higgsinos $H_{2,1}^{\pm}$. In terms of
two-component Weyl spinors, the chargino mass term in the Lagrangian
can be written as \cite{Haber}

\begin{equation} \label{eq:LagrangianX}
\mathcal{L} = -\frac{1}{2}\left(\begin{array}{cc} \psi^+ &\psi^-
\end{array}\right) \left(\begin{array}{cc} 0 &\mathcal{M}_C^T\\
\mathcal{M}_C
&0\end{array}\right) \left(\begin{array}{c} \psi^+\\
\psi^-\end{array}\right)+h.c.,
\end{equation}
which is bilinear in the fermionic fields ${\psi}_{j}^{\pm}=(-i
\widetilde W^{\pm},\widetilde H_{2,1}^{\pm})^T$.
 The chargino mass matrix $\mathcal{M}_C $ is
given by
\begin{equation} \label{eq:MassMatrixC}
\mathcal{M}_C=\left(\begin{array}{cc}M_2&{\sqrt{2}}m_Wc_\beta \\
\sqrt{2}m_Ws_\beta &|\mu|e^{i\phi_\mu}\end{array}\right).
\end{equation}
As seen from Eq.~\eqref{eq:MassMatrixC}, the matrix
$\mathcal{M}_C$ isn't symmetric; it can be diagonalized
analytically by two different unitary matrices $V$ and $U$ such
that these satisfy the relation
$U^*\mathcal{M}_CV^{-1}=\text{diag}\left\{m_{\widetilde{\chi}_{1}^{\pm}},m_{\widetilde{\chi}_{2}^{\pm}}\right\}$
with the chargino mass eigenvalues as follows:
\begin{equation} \label{eq:m_chargino}
\begin{split}
m^2_{\widetilde{\chi}_{1,2}^{+}}=&\frac{1}{2}\bigl\{M_{2}^2+{|\mu|}^2+2m^2_W\mp
\bigl[(M_2^2-{|\mu|}^2-2m^2_{W}
\cos2\beta)^2 \\
  & +8m^2_W(M_2^{2}c^2_\beta+{|\mu|}^2 s^2_\beta +
M_2|\mu|\sin2\beta \cos\phi_\mu)\bigr]^{1/2}\bigr\}.
\end{split}
\end{equation}

In this paper, we take into consideration the gaugino/Higgsino sector with
the following assumptions: We set $\phi_1=\phi_{\mu}=0$ for CP conservation.
The physical signs between $\mu$, $M_1$
and $M_2$ are relative, which could be absorbed into phases $\phi_1$
and $\phi_{\mu}$ by rearranging of fields. Therefore, $\mu$, $M_1$
and $M_2$ are chosen to be real and positive, which are usually
assumed to be related via the relation $M_1=\frac{5}{3}M_2
\tan^2\theta_W\simeq 0.5 M_2$. Using these assumptions, there appear several
scenarios for the choice of the SUSY parameters. On account of the fact
that SUSY parameters should be obtained from physical quantities,
it is also possible that we choose an alternative way to diagonalize
the mass matrix $\mathcal{M}$ by taking any two chargino
masses together with $\tan\beta$ as inputs. In this case, the two
mass parameters $M_2$ and $\mu$ can be calculated from
the chargino masses for given $tan\beta$ \cite{Choi2,Moultaka}. By
taking appropriate sums and differences of the chargino masses, one
can obtain the following solutions for $M_2$ and $\mu$:
\begin{equation} \label{eq:2M2}
M_2^2=\frac{1}{2}((m_{\widetilde{\chi}_{1}^{+}}^2+m_{\widetilde{\chi}_{2}^{+}}^2-
2m_{W}^2)\mp\sqrt{(m_{\widetilde{\chi}_{1}^{+}}^2+m_{\widetilde{\chi}_{2}^{+}}^2-
2m_{W}^2)^2-\Delta_{\pm} }),
\end{equation}
\begin{equation} \label{eq:2mu2}
|\mu|^2=\frac{1}{2}((m_{\widetilde{\chi}_{1}^{+}}^2+m_{\widetilde{\chi}_{2}^{+}}^2-
2m_{W}^2)\pm
\sqrt{(m_{\widetilde{\chi}_{1}^{+}}^2+m_{\widetilde{\chi}_{2}^{+}}^2-2m_{W}^2)^2-
\Delta_{\pm}}),
\end{equation}
with
$$
\Delta_{\pm}=4 \left[m_{\widetilde{\chi}_{1}^{+}}^2
m_{\widetilde{\chi}_{2}^{+}}^2+ m_{W}^4
cos2\phi_{\mu}sin^2{2{\beta}}\pm 2m_{W}^2 cos\phi_{\mu}
sin2\beta\times \right.
$$
$$
\left.\sqrt{m_{\widetilde{\chi}_{1}^{+}}^2
m_{\widetilde{\chi}_{2}^{+}}^2-m_{W}^4 sin^2{2{\beta}}
sin^2{\phi_{\mu}}}\right],
$$
where the lower (upper) signs correspond to the $M_2>|\mu|$ ($M_2<|\mu|$) regime.
 So, for given $tan\beta$, $\mu$ and $M_2$ , terms of the two chargino masses
$m_{\widetilde{\chi}_{1}^{+}}$ and $m_{\widetilde{\chi}_{2}^{+}}$ are obtained by
using Eqs.~\eqref{eq:2M2} and \eqref{eq:2mu2} from which one can derive four
solutions corresponding to different physical scenarios. For
$|\mu|<M_{2}$, the lightest chargino has a stronger higgsino-like
component and so it is named higgsino-like
\cite{Choi,Moultaka}. Furthermore, the solution $|\mu|>M_{2}$, corresponding to
the gaugino-like situation could be easily gotten by the
replacements as follows: $\mu\to$ sign($\mu$)$M_{2}$ and $M_{2} \to
|\mu|$ \cite{Kneur,Choi}.

\end{document}